\documentclass[aps,prb,twocolumn,showpacs,amsmath,superscriptaddress]{revtex4}
\usepackage{graphics}
\usepackage{epsfig}
\usepackage[colorlinks=true,citecolor=blue,linkcolor=red,linktocpage=true,pagebackref=false]{hyperref}
\usepackage{color,soul} 
\begin{document}

\title{
Noise of a single-electron emitter
}
\author{Michael Moskalets}
\affiliation{Department of Metal and Semiconductor Physics, NTU ``Kharkiv Polytechnic Institute", 61002 Kharkiv, Ukraine}

\date\today
\begin{abstract}
I analyze the correlation function of currents generated by the periodically driven quantum capacitor emitting single electrons and holes into the chiral waveguide. 
I compare adiabatic and non-adiabatic, transient working regimes of a single-electron emitter and find the striking difference between the correlation functions in two regimes. Quite generally for the system driven with frequency $\Omega$ the correlation function depends on two frequencies, $\omega$ and $\ell \Omega - \omega$, where $\ell$ is an integer. For the emitter driven non-adiabatically the correlation functions for different $\ell$ are similar and almost symmetric in $\omega$. While in the case of adiabatic drive the correlation functions for $\ell \ne 0$ are highly asymmetric in $\omega$ and exceed significantly the one corresponding to $\ell = 0$. Under optimal operating conditions the correlation function for odd $\ell$ is zero.  
\end{abstract}
\maketitle

\section{Introduction}
\label{intro}

\begin{figure}[t]
\begin{center}
\includegraphics[width=80mm]{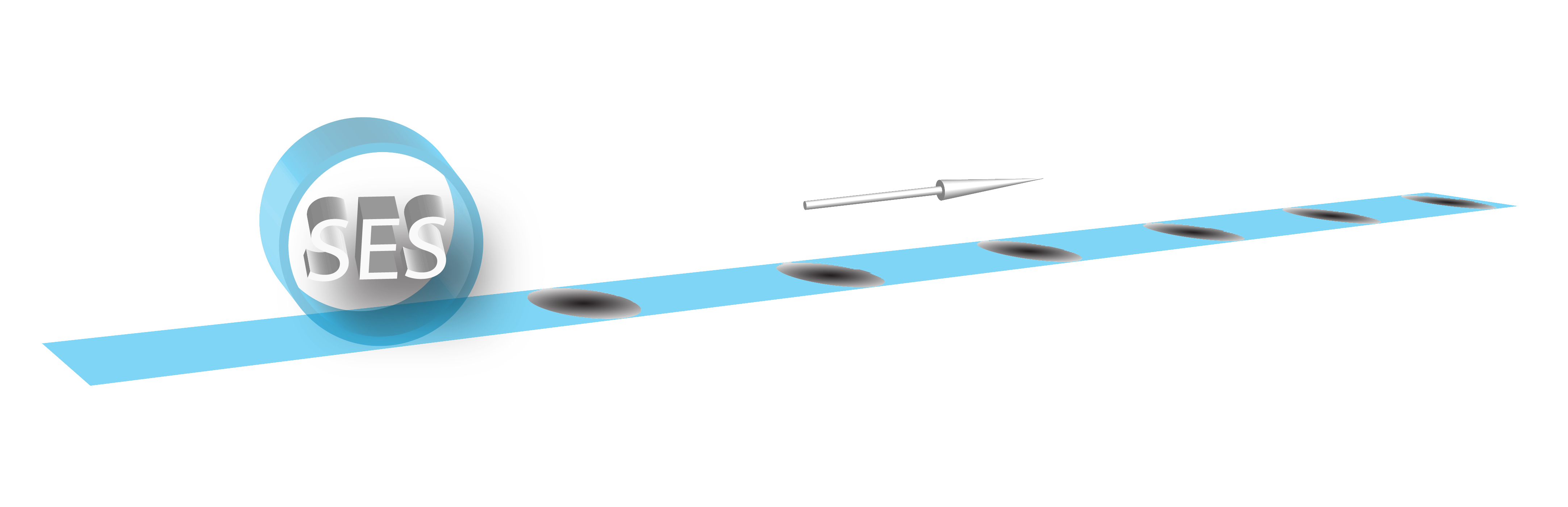}
\caption{
(Color online) Single-electron source (SES) emits the regular train of electrons or alternating electrons and holes (shown as dark areas) well separated in distance. 
The blue strip is an electron waveguide, e.g., the edge state of a two-dimensional electron gas in the integer quantum Hall effect regime. 
The arrow indicates the direction of movement of emitted particles. 
}
\label{fig1}
\end{center}
\end{figure}

Non-stationary transport in quantum systems \cite{Moskalets:2011cw} attracts recently increasing interest. 
In the dynamic regime the quantum conductors show properties revealing such fundamental features of individual carriers, which are not accessible in the stationary regime.
Thus, recent  experiments demonstrate that even a regular stream of electrons, see Fig.~\ref{fig1}, is noisy on a short time scale due to quantum fluctuations of the emission time. \cite{Mahe:2010cp}
The fluctuations of energy of emitted electrons \cite{Battista:2012vs} as well as the random waiting times between charge transfers \cite{Albert:2011fx} are also consequences of the probabilistic nature of the emission process.

The experimental implementation of high-speed single-electron sources not relying on an electron-electron interaction \cite{Blumenthal07,Feve:2007jx,Kaestner:2008gv,Fujiwara:2008gt,Dubois:2013ul} 
opens up perspectives for  the development of {\it quantum coherent electronics} (or, as it is  often referred to, {\it electron quantum optics} \cite{Grenier:2011js}), an emerging field aimed at creation \cite{Keeling:2006hq,Battista:2011jb,Vanevic:2012bg,Dubois:2012us,Fletcher:2012te}, manipulation \cite{Olkhovskaya:2008en,Kataoka:2011fu,Juergens:2011gu,Sherkunov:2012dg,Jonckheere:2012cu,Bocquillon:2013dp}, and transportation \cite{Hermelin:2011du,McNeil:2011ex} of single to few particles wave-packets in the condensed matter. 
To characterize the coherence properties of a single-electrons state the few protocols were already suggested. \cite{Grenier:2011dv,Grenier:2011js,Haack:2011em,Haack:2012wa} 

One of the most important questions for quantum information processing applications \cite{Bennett:2000kl} is whether the emitted state is a genuine single-particle state. 
In quantum optics the single-photon state is probed via the (no-)coincidence measurement. \cite{Lounis:2005ex} 
In mesoscopics the current noise measurement is more suitable for this purpose. 
It is known that the zero-frequency noise vanishes if the stream of electrons is regular. \cite{Blanter:2000wi} 
Indeed, the measurements \cite{Maire:2008hx}  show a significant reduction of the low-frequency noise power in the quantized emission regime, when the source emits the same number of electrons, in particular one electron, during each working period.  
The direct counting of the number of particles emitted in each cycle was demonstrated by measuring the partition noise of the beam-splitter \cite{Bocquillon:2012if} and with the help of a charge detector \cite{Fricke:2012vk}. 

As the experiment reveals, the noise spectrum provides an accurate characterization of a single-electron emission regime. \cite{Mahe:2010cp} 
If the source works as a single-electron emitter then the excess finite-frequency noise becomes minimal though not zero.  
To shed light onto the nature of this fundamental noise limit, the semiclassical model of a single-electron emitter, consisting in a particle periodically attempting to leave the cavity was put forward. \cite{Mahe:2010cp,Albert:2010co,Parmentier:2012ed} 

This model,  shows clearly that the uncertainty in the emission time puts the lower limit to the excess finite-frequency noise. 
This  limiting noise is referred to as {\it the phase noise}.  
The phase noise limit was achieved in  Ref.~\onlinecite{Mahe:2010cp} for the source driven by the pulsed potential and thus emitting electrons non-adiabatically.

Our aim here is to analyze the entire spectrum of the noise generated by the  single-particle source. 
In addition to the noise present already in the stationary case \cite{Buttiker:1993wh} and modified by the working single-particle source, -  this modification is referred to as {\it the excess noise}, -  there is also the photon-induced noise completely absent in the stationary case. 
To calculate the noise spectrum we use the Floquet scattering matrix theory \cite{Moskalets:2002hu} for both adiabatic and non-adiabatic emission \cite{Moskalets:2013dk}. 
For non-adiabatic optimal working conditions our analytical results for the phase noise are in agreement with both the solutions of the semiclassical model and the results of numerical calculations based on the Floquet approach to photon assisted noise, which were contrasted in Ref.~\onlinecite{Parmentier:2012ed}.  
As we will show the phase noise in the adiabatic regime exhibits similar behavior: At low measurement frequencies $\omega$ the phase noise grows quadratically while with increasing frequency it gets suppressed. 
The photon-induced noise generated in the non-adiabatic regime resembles the phase noise. 
In contrast, the photon-induced noise in the adiabatic regime shows completely different behavior: It exists within the finite frequency   window, is asymmetric in frequency, and far exceeds the phase noise in magnitude. 
This is one more example illustrating a striking difference between the quantum states of electrons emitted adiabatically and non-adiabatically, see also Ref.~\onlinecite{Moskalets:2013dk}.     

The paper is organized as follows. 
In Sec.~\ref{ffn} we use the Floquet scattering matrix formalism to derive the correlation function for currents generated by the quantum capacitor attached to the chiral electron waveguide. 
The scattering amplitude of a periodically driven capacitor is presented and analyzed in Sec.~\ref{Floquet}. 
The spectrum of the current correlation function in adiabatic and non-adiabatic regimes is presented in Secs.~\ref{en} and \ref{pin} for the excess noise and the photon-induced noise, respectively. 
We conclude in Sec.~\ref{concl}.
Some details of calculations are given in appendices.

\section{Finite-frequency noise of a capacitor attached to the Hall bar}   
\label{ffn}

\begin{figure}[b]
\begin{center}
\includegraphics[width=80mm]{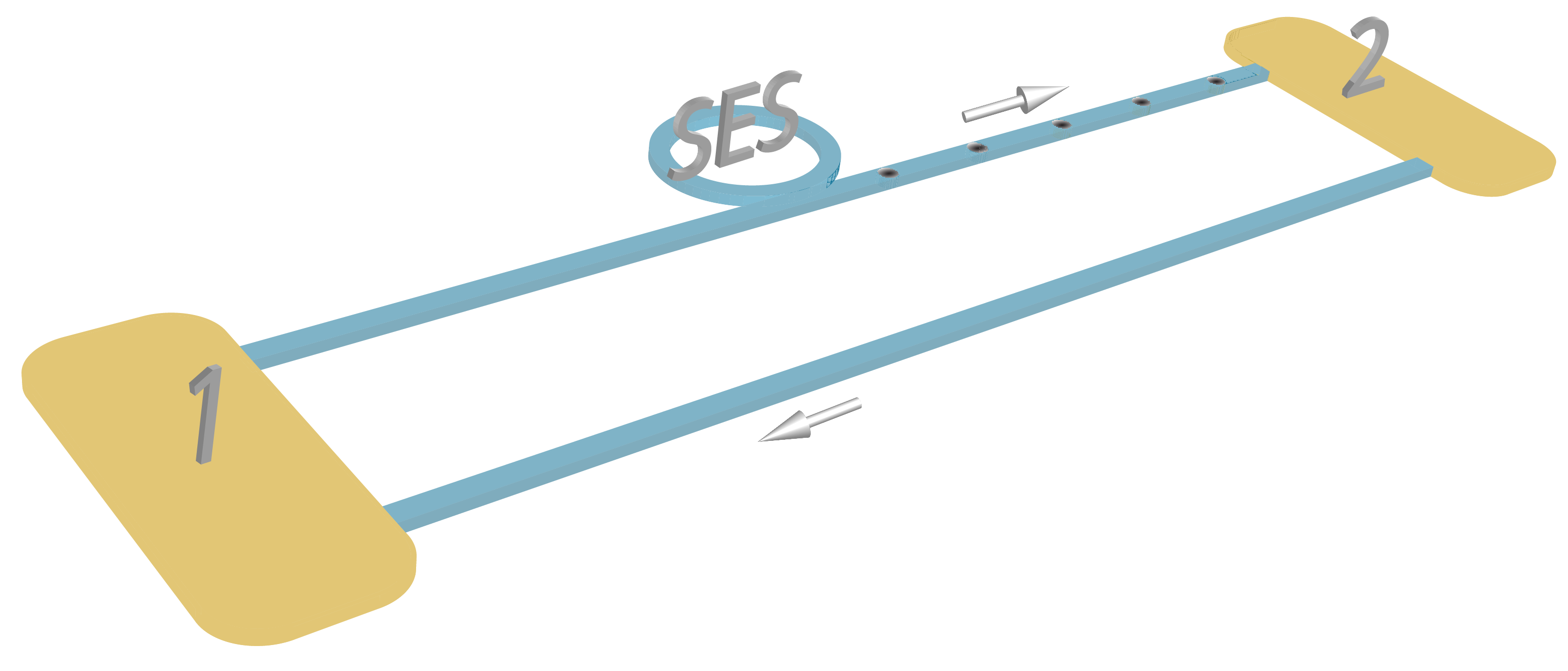}
\caption{(Color online) Single-electron source (SES), a circular edge state driven by the periodic potential of the top-gate (not shown), emits electrons and holes (shown as dark areas) into one of the edge states of the Hall bar. The arrows indicate the direction of movement of emitted particles as well as of electrons of the underlying Fermi sea in the edge states (shown as blue strips). $1$ and $2$ are metallic contacts. The contact $1$ is grounded. The current and its fluctuations are measured at the contact $2$ where the emitted particles come to.    
}
\label{fig2}
\end{center}
\end{figure}

The cartoon of the set-up of  Ref.~\onlinecite{Mahe:2010cp} is shown in Fig.~\ref{fig2}. 
The cavity, a circular edge state, driven by the periodic potential of the top-gate, $U(t) = U(t + 2\pi/\Omega)$, emits electrons and holes into the chiral edge state of the Hall bar.
The working principle of this single-electron source is as follows.
The sample is made in a two-dimensional electron gas in the quantum Hall effect regime \cite{Klitzing:1980wf}. 
In this regime the bulk of the sample is not conducting. 
However there exist conducting states along the edges of a sample. 
Electrons propagate within edge states without backscattering. \cite{Halperin:1982tb,Buttiker:1988vk}  
Therefore, these states are referred to as the chiral edge states.
The emitter is a small cavity with quantized electron spectrum which is side-coupled to the chiral electron waveguide with the chemical potential $\mu$. 
The periodic potential $U(t)$ drives cavity's levels up and down. 
When ${\rm e}U(t)$ increases and some cavity's level crosses $\mu$ an electron is emitted to the waveguide, since in the waveguide only the states with energy larger than $\mu$ are empty (for simplicity we consider zero temperature).
In the reverse process an electron enters the cavity, a hole is emitted to the waveguide. 
In total during the period no net charge is emitted to the waveguide. 
Therefore, such a cavity can be referred to as a quantum capacitor. \cite{Buttiker:1993wh,Gabelli:2006eg}

The essential feature of the measurement set-up is the presence of two independent metallic contacts, $1$ and $2$ playing the role of the source and drain, respectively, for an electron waveguide. 
These contacts have the same chemical potential $\mu$, the same temperature, $T$,  and  thus the same Fermi distribution function $f(E)$.
Only one of contacts, namely the contact $2$, is used to measure the current $I(t)$, associated with particles emitted by the capacitor, and its fluctuations. 
Importantly, since the current flowing into and out of the contact $2$ are not correlated, it is possible to extract the contribution to current fluctuations solely due to emitted particles. \cite{Parmentier:2012ed}

\subsection{Current correlation function}

The current fluctuations at contact $2$ we characterize with the help of the current correlation function, \cite{Blanter:2000wi} 

\begin{equation}
{\rm P}_{22}(\omega_{1},\omega_{2}) = \frac{1}{2} \left\langle \delta \hat I_{2}(\omega_1) \delta \hat I_{2}(\omega_2) + \delta \hat I_{2}(\omega_2)  \delta \hat I_{2}(\omega_1) \right\rangle .
\label{Eq2_3_09}
\end{equation}
\ \\
\noindent
Here the angle brackets $\left\langle \dots \right\rangle$ stands for quantum-statistical average over the equilibrium state of electron reservoirs;  
$\delta \hat I_{2}(\omega) = \hat I_{2}(\omega) - \left\langle \hat I_{2}(\omega) \right\rangle$ is an operator of current fluctuations in the contact $2$. 
The current operator, $\hat I_{\alpha}(2)$, is expressed in terms of the second quantization creation, $\hat b^{\dag}_{2}$, and annihilation, $\hat b_{2}$, operators for particles moving in the wave guide and entering the reservoir $2$ and the operators $\hat a_{2}^{\dag}$ and $\hat a_{2}$ for particles leaving the reservoir $2$. 
If the relevant energy scales (such as the voltage applied, the temperature difference, the energy quantum $\hbar\Omega$, etc.) characterizing the non-equilibrium state are small compared to the Fermi energy $\mu$, then the current operator reads, \cite{Buttiker:1992vr}

\begin{eqnarray}
\hat I_{2}(\omega) &=& {\rm e}\!\! \int\limits_{0}^{\infty} \!\! dE \! \left\{ \hat b_{2}^{\dagger}(E) \hat b_{2}(E+\hbar\omega) - \hat a_{2}^{\dagger}(E) a_{2}(E + \hbar\omega) \right\} .
\nonumber \\
\label{Eq2_3_10} 
\end{eqnarray}
\ \\
\noindent
In the case shown in Fig.~\ref{fig2}, the particles entering the contact $2$ were originally emanated by the contact $1$ and subsequently scattered by the capacitor. 
Since the capacitor is driven by the periodic potential with frequency $\Omega$, an electron scattered off the capacitor can gain or loss one or several energy quantum $\hbar\Omega$. 
Such a scattering is described by the Floquet scattering matrix, which we will denote as $\hat S_{F}$.   
Therefore, we write, \cite{Moskalets:2002hu}

\begin{eqnarray}
 \hat b_{2}(E) = \sum\limits_{n=-\infty}^{\infty} S_{F}\left( E, E_{n} \right) \hat a_{1}\left( E_{n} \right)\,,
\label{df}
\end{eqnarray}
\ \\
\noindent
where $E_{n} = E + n\hbar\Omega$. 
The element $S_{F}\left( E, E_{n} \right)$ is a forward-scattering photon-assisted amplitude for electrons    propagating in the waveguide and passing through the place where the capacitor is attached to. 
We stress, the initial equations are written in terms of operators for electrons propagating in the waveguide, however the final equations can be interpreted as such which describe the properties of particles emitted by the source attached to the waveguide.  

The electron reservoirs, contacts $1$ and $2$, are assumed to be in equilibrium.
Therefore, the operators, $\hat a_{\alpha}$, $\alpha = 1,2$, for particles emanating by these contacts satisfy the following relations for equilibrium fermions: 
\begin{eqnarray}
\hat a_{\alpha}^{\dagger}(E)\, \hat a_{\beta}(E^{\prime}) + \hat a_{\beta}(E^{\prime})\,  \hat a_{\alpha}^{\dagger}(E) &=& \delta_{\alpha,\beta}\, \delta(E - E^{\prime})\,, 
\nonumber \\
\left\langle \hat a_{\alpha}^{\dag}(E) a_{\alpha}(E^{\prime})  \right\rangle &=& f(E) \delta(E-E^{\prime})\,.
\label{acav} 
\end{eqnarray}
\ \\
\noindent  
Here $\delta_{\alpha,\beta}$ is the Kronecker symbol and $\delta(E - E^{\prime})$ is the Dirac delta function. $f(E)$ is the Fermi distribution function the same for both contacts, $f_{\alpha}(E) = f(E)$, $\alpha = 1, 2$. 

For the reference purposes we give a time-dependent current, $I_{2}(t) = \int d\omega/(2\pi)\, e^{- i \omega t}\left\langle  \hat I_{2}(\omega)  \right\rangle$, generated in the contact $2$.  
Substituting Eq.~(\ref{df}) into Eq.~(\ref{Eq2_3_10}) and calculating quantum-statistical averaging with the help of Eq.~(\ref{acav}) we obtain, \cite{Splettstoesser:2008gc} 

\begin{eqnarray}
I_{2}(t) = \sum\limits_{\ell = - \infty}^{\infty} e^{- \ell \Omega t} \frac{\rm e }{h } \int\limits_{0}^{\infty} dE
\nonumber \\
\times
 \sum\limits_{n = -\infty}^{\infty} S_{F}^{*}\left( E_{n}, E \right) S_{F}\left( E_{n+\ell}, E \right) \left\{ f(E) - f(E_{n})\right\} 
\,.
\label{i2t}
\end{eqnarray}
\ \\
\noindent

Now we calculate the current correlation function. 
Substituting Eq.~(\ref{Eq2_3_10}) into Eq.~(\ref{Eq2_3_09}) and calculating quantum-statistical averaging as before we  arrive at the following general equation for the current correlation function, \cite{Moskalets:2007dl,Moskalets:2011jx}

\begin{subequations}
\label{001}
\begin{eqnarray}
{\rm P}_{22}(\omega_{1},\omega_{2}) = \sum\limits_{\ell = -\infty}^{\infty} 2\pi \delta\left( \omega_{1} + \omega_{2} - \ell \Omega \right) {\cal P}_{\ell}(\omega_{1}) \,,
\label{001A}
\end{eqnarray}
\ \\
\noindent
with the noise power 

\begin{eqnarray}
{\cal P}_{\ell }(\omega)  = \frac{{\rm e}^{2}  }{h} \int\limits_0^\infty  dE\,\bigg\{ \delta _{l0}\, F \left( {E,E + \hbar \omega } \right)
\nonumber \\
+  \sum\limits_{n,m,p =  - \infty }^\infty   F (E_{ \ell + n} ,E_m  + \hbar \omega ) 
\nonumber \\
\nonumber \\
\times S_{F}^* \left( E,E_{ \ell + n} \right)  S_{F} \left( E + \hbar \omega,E_m  + \hbar \omega  \right)\nonumber \\
\nonumber \\
\times S_{F} \left( E_{ \ell + p} ,E_{ \ell + n}  \right) S_{F }^* \left( E_p  + \hbar \omega,E_m  + \hbar \omega  \right)\!\! \bigg\} , 
\label{001B}
\end{eqnarray}
\ \\
\noindent
and the symmetrized combination of the Fermi functions

\begin{equation}
F(E_{1},E_{2}) = \frac{ f(E_{1})\left [ 1 - f(E_{2}) \right] + f(E_{2})\left [ 1 - f(E_{1}) \right] }{2} \,.
\label{002}
\end{equation}
\end{subequations}
\ \\
\noindent
Notice with small $\omega_{1}$ and $\omega_{2}$ we denote frequencies  which the correlation function is measured at, while with a large $\Omega$ we denote the frequency of the potential driving the source. 
Within the Floquet scattering matrix formalism the component of the noise power with $\ell = 0$  was addressed in Ref.~\onlinecite{Pedersen:1998uc}. 

One can check that the noise power satisfies the following symmetries, \cite{Moskalets:2009dk}

\begin{eqnarray}
{\cal P}_{\ell }(\omega) = {\cal P}_{\ell }(\ell \Omega - \omega) \,,
\nonumber \\
{\cal P}_{\ell }(\omega) 
= 
\left\{ 
{\cal P}_{-\ell }(-\omega)  
\right\}^{*} 
\,.
\label{45-1} 
\end{eqnarray}

\subsection{Excess noise}

The central component, ${\ell} = 0$, of the correlation function, Eq.~(\ref{001}), is not zero even in equilibrium, when the driving potential is switched off. 
This is due to the quantum noise. \cite{Gardiner:2000wq,Clerk:2010dh} 
To assess the contribution due to the working SES only, the difference between the correlation functions (corresponding to $\ell=0$) with the SES on and off was measured. \cite{Mahe:2010cp} 
This difference is referred to as the excess noise.  
We will denote it with the superscript ``$ex$''
After the simple algebra we get from Eq.~(\ref{001B}),

\begin{eqnarray}
{\cal P}^{ex}(\omega) = \frac{{\rm e}^{2}  }{h} \int\limits_0^\infty  dE  \sum\limits_{m =  - \infty }^\infty    F ( {E ,E_{m}  + \hbar \omega } )  
 \nonumber \\ 
 \times \left\{ \left | \Pi_{m}(E,\omega) \right |^{2}  - \delta_{m,0}  \right\} ,
\label{33}
\end{eqnarray} 
where 
\ \\
\noindent
\begin{equation}
\Pi_{m}(E,\omega) =   \sum_{q=-\infty}^{\infty} S_{F} \left( E_{q},E \right) S_{F}^* \left( E_{q}+ \hbar \omega ,E_{m} + \hbar \omega \right) .
\label{34}
\end{equation}
\ \\
\noindent
The similar equation was found in  Ref.~\onlinecite{Parmentier:2012ed}.

\subsection{Photon-induced noise}

In the stationary case the currents at $\omega_{1} \ne - \omega_{2}$ are not correlated and  the corresponding terms of the current correlation function, Eq.~(\ref{001A}), all are zero.
In contrast, in the dynamic case the currents at $\omega_{1}$ and $\omega_{2} = \ell \Omega - \omega_{1}$ are correlated \cite{Moskalets:2007dl,Gabelli:2008eo} due to the ability of a dynamic scatterer to provide (emit or absorb) photons with energy $\hbar\Omega$. 
Therefore, the side components, ${\ell} \ne 0$, of the correlation function, Eq.~(\ref{001}), appear. 
We name the corresponding noise as {\it the photon-induced noise} (PIN). 
The spectrum of the PIN power reads, 

\begin{eqnarray}
{\cal P}_{\ell }^{PIN}(\omega)  &=& \frac{{\rm e}^{2}  }{h}  \int\limits_0^\infty  dE   \sum\limits_{m =  - \infty }^\infty   F \left( {E ,E_m  + \hbar \omega } \right)
\nonumber \\
&& \times  \Pi_{m}^{\ell}(E,\omega) \Pi_{m}^{0 *}(E,\omega)  ,
\label{43}
 \end{eqnarray}
where

\begin{equation}
\Pi_{m}^{\ell}(E,\omega)  = \!\! \sum\limits_{p =  - \infty }^\infty \!\! S_{F}^{} \left( {E_{p + \ell} ,E } \right) S_{F }^* \left( {E_{p }  + \hbar \omega,E_m  + \hbar \omega } \right) .
\label{n34} 
\end{equation}

To proceed with calculations we need to specify the model describing an electron source and to calculate the Floquet scattering matrix, $\hat S_{F}$, of the source. 

\section{Floquet scattering matrix of a driven capacitor}
\label{Floquet}

In order to calculate the Floquet scattering matrix elements $S_{F}(E_{n}, E)$ entering  Eqs.~(\ref{34}) and (\ref{n34})  we use the following model: \cite{Pretre:1996uw,Gabelli:2006eg}
Electrons in the cavity propagate along a single-channel chiral state of length $L$, which is coupled with the help of a quantum point contact (QPC) to a single-channel linear edge state playing the role of an electron waveguide.    
The QPC has energy-independent reflection, $r$, and transmission, $\bar t$, amplitudes.  
If the electron spectrum can be linearized, $k(E) \approx k(\mu) + (E - \mu) \partial k/\partial E$, then this model admits an exact  solution: \cite{Moskalets:2008fz} 

\begin{equation}
S_{F}\left(E, E_{n}\right) = S_{out, -n}(E)  \equiv \int\limits_{0}^{\cal T} \frac{dt}{\cal T} e^{-in\Omega t} S_{out}(E,t) \,.
\label{03}
\end{equation}
with
\begin{eqnarray}
S_{out}(E,t) &=& \sum\limits_{q=0}^{\infty}  S^{(q)}(t)\,, 
\nonumber \\
S^{(0)} &=& r\,, \quad S^{(q>0)}(t) = \bar t^2\, r^{q-1}\,e^{iqkL}\, e^{-i\Phi_{q}(t)}\,, 
\nonumber \\
\Phi_q(t) &=& \frac{{\rm e} }{\hbar}\int\limits_{t}^{t+q\tau} dt^\prime U(t^\prime) \,. 
\label{04}
\end{eqnarray}
\ \\
\noindent
Here $\tau$ is the time of one turn around the cavity. 
The physical meaning of the amplitude $S_{out}$ is discussed in Ref.~\onlinecite{Moskalets:2008ii}. 

Given above general solution is not restricted to any particular amplitude and/or time--dependence of the driving potential $U(t)$. 
However the scattering amplitude becomes especially simple in two limiting cases: (i) When the potential changes slow, adiabatically and (ii) when the time-dependent potential consists of the sequence of pulses. In the later case the potential changes non-adiabatically. \cite{Moskalets:2013dk}
The time scale differentiating slow and fast variations is put by the dwell time $\tau_{D}$, the time necessary for an electron to leave the cavity starting from the time when the unoccupied states outside become available. 
Below we are interested in the case when the time interval between electron ($t_{-}$) and hole ($t_{+}$) emission is long enough such that the particles are definitely emitted,

\begin{eqnarray}
|t_{-} - t_{+}| \gg \tau_{D} \,. 
\label{08}
\end{eqnarray}
\ \\
\noindent
This regime is referred to as {\it the quantized emission regime}.  
For the discussion of a short-period case see Refs.~\onlinecite{Mahe:2010cp,Albert:2010co,Parmentier:2012ed}.

\subsection{Adiabatic emission}

When the potential changes slow \cite{Moskalets:2002hu}, we can keep $U(t^{\prime})$ constant while integrating over $t^{\prime}$ in Eq.~(\ref{04}). 
Then the scattering amplitude becomes the frozen scattering amplitude, $S_{out}(E,t) = S(E,t)$, of a Fabry-Perot type, \cite{Moskalets:2008fz}

\begin{equation}
S(E,t) = - e^{ i(\phi(E,t) + \theta_{r}) } \frac{ 1 -   \sqrt{R} e^{-i\phi(E,t)} }{ 1 - \sqrt{R} e^{i\phi(E,t)} } \,,
\label{06}
\end{equation}
\ \\
\noindent
where $\theta_{r}$ and $R$ are the phase and the square absolute value of the QPC's reflection amplitude, $r = \sqrt{R}\exp(i\theta_{r})$. 
The time-dependent phase acquired by an electron during one turn around the cavity is 

\begin{eqnarray}
 \phi(E,t) &=& \theta_{r} + k_{\mu}L + \frac{2\pi \left( E-\mu \right)   }{ \hbar/\tau } -  \frac{ {\rm e}U(t)  }{ \hbar/\tau } \,, 
 \label{05} 
\end{eqnarray}
\ \\
\noindent
with $\tau = \Delta/h$ a duration of one turn and $\Delta$ the level spacing in the cavity. 

Here we are concerned with the case when the QPC's transmission is small, $\bar T = 1 - R \ll 1$ and, correspondingly, the width $\bar \delta$ of the levels in the cavity is small compared to the level spacing. For the equidistant spectrum it is $\bar\delta = \bar T \Delta/(4\pi)$. 
In this case the levels in the cavity can be modeled as the Breit-Wigner resonances   \cite{Breit:1936ud} and the scattering amplitude can be represented as follows,

\begin{eqnarray}
S(E,t) = e^{i\theta_{r}} \sum\limits_{n} \frac{t - t_{\mp}^{(n)}(E) \pm i \Gamma_{\tau}^{(n)}(E) }{t - t_{\mp}^{(n)}(E) \mp i \Gamma_{\tau}^{(n)}(E) } \,,
\label{06-1}
\end{eqnarray}
\ \\
\noindent
where $n$ numbers the quantum levels in the cavity.
The upper (lower) sign is for the case when the levels go up (down) under the  action of the potential $U(t)$. 
The time $t_{\mp}^{(n)}(E)$ is a time when the $n$th quantum level has an energy $E$, and $\Gamma_{\tau}^{(n)}(E)$ is a parameter, which can be interpreted as the time interval during which the level of width $\bar\delta$ crosses the energy $E$. 
Below we consider regime when only one level, $n=0$, crosses the Fermi level. 
For the sake of short notation we will drop the level number superscript and use, $t_{\mp} = t_{\mp}^{0}$ and $\Gamma_{\tau} = \Gamma_{\tau}^{0}$. 

\begin{figure}[t]
\begin{center}
\includegraphics[width=80mm]{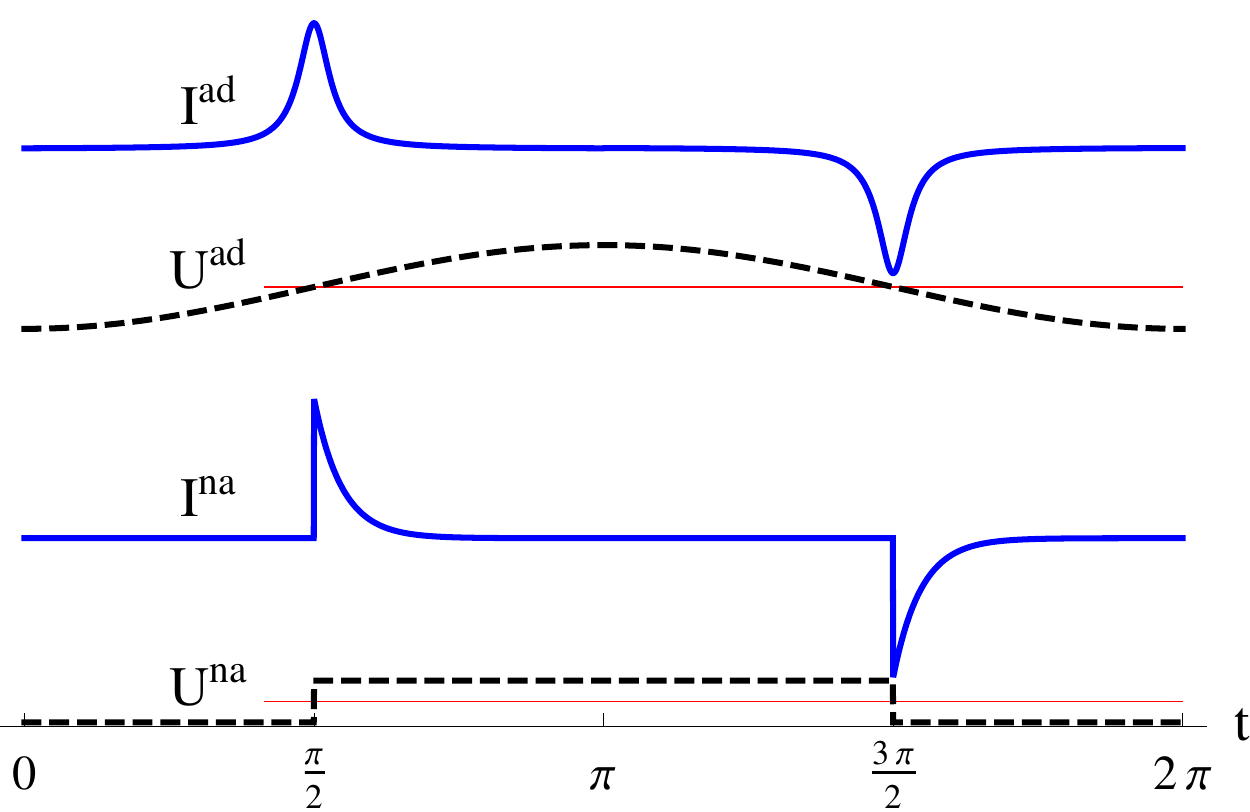}
\caption{(Color online) 
Time-dependent potential $U$ (black dashed line) driving the capacitor and the corresponding generated time-dependent current $I$ (blue solid line). The upper indices stand for adiabatic ``$ad$'' and non-adiabatic ``$na$'' regimes. The red thin line indicates the position of the Fermi level. When the cavity's level rises above the Fermi level an electron is emitted at time $t_{-} = \pi/2$ (upward current pulse). When the cavity's level sinks below the Fermi level, a hole is emitted at time $t_{+} = 3\pi/2$ (downward current pulse). Potential $U$ and current $I$ are given in arbitrary units. Time $t$ is given in units of $1/\Omega$. 
}
\label{fig3}
\end{center}
\end{figure}

At the adiabatic drive the Floquet scattering matrix element can be expressed (to the leading order in small $\Omega$) as the Fourier coefficient of the frozen scattering amplitude, $S_{F}(E_{n},E) \approx S_{n}(E)$. 
The use of this approximation in Eq.~(\ref{i2t}) results in the following current (at zero temperature): $I_{2}(t) = i{\rm e}/(2\pi) S^{*}(\mu,t) \partial S(\mu,t)/\partial t$. \cite{Buttiker:2006wt}
The frozen scattering amplitude given in Eq.~(\ref{06-1}) leads to the current consisting of the Lorentzian pulses.\cite{Olkhovskaya:2008en,Keeling:2008ft} 
If only one level crosses the Fermi level during the period, the current is (see Fig.~\ref{fig3}) 

\begin{eqnarray}
I^{ad}(t) = 
\frac{{\rm e} \Gamma_{\tau}/\pi }{ \left( t-t_{-} \right)^{2} +\Gamma_{\tau}^{2} } 
-
\frac{{\rm e} \Gamma_{\tau}/\pi }{ \left( t-t_{+} \right)^{2} +\Gamma_{\tau}^{2} } 
\,.
\label{itad}
\end{eqnarray}
\ \\
\noindent
Here the superscript ``$ad$'' stands for the adiabatic regime; $t_{-}$ ($t_{+}$) is the time of an electron (a hole) emission. 
In the above equation the current is given for the single period, $0 < t < {\cal T}$ only. 
For other times it should be extended periodically, $I^{ad}(t + n{\cal T} ) = I^{ad}(t)$, where $n$ is an integer.  

Given the current pulse shape we can formulate the adiabaticity condition more precisely,\cite{Moskalets:2013dk}

\begin{eqnarray}
\Gamma_{\tau} \gg \tau_{D} \,.
\label{adcond}
\end{eqnarray}
\ \\
\noindent
That is, the time scale $\Gamma_{\tau}$ over which the current varies has to be large compared to the time during which an electron escapes the cavity, the dwell time $\tau_{D}$. 
Note, in the adiabatic regime the condition guaranteeing that the particle is emitted is more strict than Eq.~(\ref{08}). 
It is the following, 

\begin{eqnarray}
|t_{-} - t_{+}| \gg \Gamma_{\tau} \,.
\label{qead}
\end{eqnarray}

\subsection{Non-adiabatic emission: Periodic step potential}
\label{step}

Now we analyze the case when the potential $U(t)$ changes fast compared to $\tau_{D}$. 
As an example we choose the case when $U(t)$ changes in a step-like manner between the values $U_{0}$ and $U_{1}$ during the period ${\cal T}$: 

\begin{equation}
U(t) = \left\{
\begin{array}{lc}
U_{0} \,,  &-{\cal T}/2 < t <  t_{-}\,,\\
 &  \\
U_{1} \,,  &t_{-} < t <  t_{+}\,,\\
 &  \\
U_{0} \,,  & t_{+} < t <  {\cal T}/2\,.
\end{array}
\right.
\label{07} 
\end{equation} 
\ \\
\noindent
Let us consider only the first potential jump and assume that before $t < t_{-}$ an electrons system was in equilibrium corresponding to the cavity's potential $U = U_{0}$.  
When the potential changes abruptly at $t=t_{-}$ the system is driven into a non-equilibrium state. 
It takes a time of order $\tau_{D}$ for the system to adjust to a new value of the potential $U(t)$ and to equilibrate. 
During this equilibration transient process an electron (if ${\rm e}(U_{1} - U_{0}) > 0$) or a hole (if ${\rm e}(U_{1} - U_{0}) < 0$)  is emitted.  
After this time, $t-t_{-} \gg \tau_{D}$, the system is in a new equilibrium state, which corresponds to the cavity's potential $U = U_{1}$.  
Thus, if the time-delay between the two potential steps is long enough, see Eq.~(\ref{08}), then the electron system has enough time to react to each potential step independently. 
Therefore, the Floquet scattering matrix corresponding to the periodic step potential, Eq.~(\ref{07}), can be expressed in terms of the Floquet scattering matrices for each potential step. 

Let us consider the first potential step (for the safe of short notation we put $t_{-} = 0$),

\begin{equation}
U^{s}(t) = \left\{
\begin{array}{ll}
U_{0} \,,  &t <  0\,,\\
 &  \\
U_{1}\,, &  t > 0 \,.
\end{array}
\right.
\label{09} 
\end{equation}
\ \\
\noindent
The superscript ``$s$'' indicates a single step case. 
The time-dependent phase $\Phi_{q}(t)$,  Eq.~(\ref{04}), can be calculated as follows,

\begin{equation}
\Phi_{q}^{s}(t) = \left\{
\begin{array}{lc}
2\pi  \frac{{\rm e}U_{0}}{\Delta} q\,, &  t < - q\tau \,,\\
\ \\
2\pi \frac{{\rm e} \delta U  }{\Delta} \frac{t}{\tau} + 2\pi  \frac{{\rm e} U_{1}}{\Delta} q\,, & -q\tau <  t < 0 \,, \\
\ \\
2\pi  \frac{{\rm e}U_{1}}{\Delta} q\,, & t > 0\,.\\
\end{array}
\right.
\label{10} 
\end{equation}
\ \\
\noindent
Here $\delta U = U_{1} - U_{0}$.
In order to sum up over $q$ in Eq.~(\ref{04}) we do the following. 
For any $t > 0$ we use $U = U_{1}$ $\forall q$.
If $ -(N+1)\tau < t < -N\tau$ we use $U = U_{0}$ for $q \leq N$ and take into account the potential change for $q > N$.
Then we calculate the time-dependent scattering amplitude $S_{out}(E,t)$, Eq.~(\ref{04}), as follows:\cite{Moskalets:2013dk}

\begin{equation}
S_{out}^{s}(E,t) = \left\{
\begin{array}{ll}
S(E,U_{0}) + \delta S(E,t), & t < 0 ,\\
\ \\
S(E,U_{1}) \,, & t > 0\,.
\end{array}
\right.
\label{11} 
\end{equation}
\ \\
\noindent
Here $S(E,U_{j})$, $j=0,1$, is the frozen scattering amplitude, Eq.~(\ref{06}), of the cavity having the potential $U(t) = U_{j}$. 
The time-dependent part of the scattering amplitude $\delta S(E,t)$ can be cast into the relatively simple form separately for each interval of duration $\tau$: $\delta S(E,t)  = \delta S_{N}(E,t)$ for $-(N+1)\tau < t < - N\tau$ where $N$ is an integer  and 

\begin{eqnarray}
\delta S_{N}(E,t) = - e^{ i\theta_{r} }\, \bar T\, \sqrt{R^{N}} 
\nonumber \\
 \times  \left\{ e^{-2\pi i \frac{e\delta U}{\Delta}\frac{t}{\tau} } \frac{ e^{i(N+1)\phi_{1}} }{1 - \sqrt{R} e^{i\phi_{1}} } -   \frac{ e^{i(N+1)\phi_{0}} }{1 - \sqrt{R} e^{i\phi_{0}} } \right\}  . 
\label{12} 
\end{eqnarray}
\ \\
\noindent
Here $R$ is the reflection probability of the QPC, $\bar T = 1 - R$ is the transmission probability of the QPC, and $\phi_{j}$, $j= 0,\, 1$, is the phase defined in Eq.~(\ref{05}) with $U(t)$ replaced by $U_{j}$. 

Now we turn back to the periodic step potential, Eq.~(\ref{07}), and under the condition set out in Eq.~(\ref{08}) we calculate the corresponding scattering amplitude.  
To make equation more compact, it is convenient to take as a period the following time interval, $t \in ( t_{+} - {\cal T} ; t_{+})$.
Then we can write  

\begin{equation}
S_{out} = \left\{
\begin{array}{ll}
S(E,U_{0}) + \delta S(E,t-t_{-})\,, &   t_{+} - {\cal T}  <  t < t_{-} \,,\\
\ \\
S(E,U_{1}) + \delta \tilde S(E,t-t_{+}) \,, & t_{-} < t < t_{+} \,.
\end{array}
\right.
\label{13} 
\end{equation}
\ \\
\noindent
Here $\delta\tilde S$ is the same as $\delta S$ but with $U_{0}$ and $U_{1}$ being interchanged. 
Finally, to get the Floquet scattering matrix elements, Eq.~(\ref{03}) we have to calculate the Fourier coefficient for the function given above.  
It can be simply done in the general case. 

However below we restrict ourselves to the experimentally important case when the potential is changed by exactly one level spacing, ${\rm e}\delta U = \Delta$, and the potential $U_{0}$ is chosen in such a way that the Fermi level is positioned exactly in the middle between the two subsequent levels in the cavity. 
These working conditions are referred to as {\it the optimal working conditions}. \cite{Feve:2007jx} 
At the non-optimal operating conditions the noise is enhanced indicating that the source works not as a single electron emitter.\cite{Parmentier:2012ed,Jonckheere:2012be}

\subsubsection{Optimal operating conditions}
\label{optimal}

For the optimal operation conditions  $\phi_{1} = \phi_{2}\equiv \phi^{opt}$ with 

\begin{eqnarray}
\phi^{opt}(E) &=& \pi + 2\pi \left( E - \mu \right)/\Delta \,, 
 \label{14} 
\end{eqnarray}
and
\begin{eqnarray}
\delta S_{N}^{opt} = - e^{ i\theta_{r} } \bar T \sqrt{R^{N}} \frac{ e^{i(N+1)\phi^{opt}(E) } \left\{ e^{-2\pi i \frac{t}{\tau} }  -   1 \right\}  }{1 - \sqrt{R} e^{i\phi^{opt}(E) } } .
\label{15} 
\end{eqnarray}
\ \\
\noindent
To calculate the Fourier coefficients of the function $S_{out}(E,t)$, Eq.~(\ref{13}), we represent the integral over time in Eq.~(\ref{03}) as the sum of integrals over the time interval of duration $\tau$. 
Under the condition set out in Eq.~(\ref{08}) this sum contains the infinite number of terms. 
Thus, substituting Eq.~(\ref{15}) into Eq.~(\ref{13}) and subsequently into Eq.~(\ref{03}) we find after a little algebra the Floquet scattering matrix elements for the optimal operating conditions, 

\begin{eqnarray}
S_{F}^{opt}\left(E, E_{n}\right) = S(E)  \delta_{n,0} \!-\!   A_{n}\left\{ \! \frac{ e^{-i n \Omega t_{-}} }{1 +  \frac{n\hbar\Omega }{\Delta }  } - \frac{ e^{-i n \Omega t_{+}} }{1 -  \frac{n\hbar\Omega }{\Delta } }  \! \right\}  , 
\nonumber \\
\label{16} 
\end{eqnarray} 
where $E_{n} = E + n\hbar\Omega$.  
$S(E)$ is  the stationary scattering amplitude at $U=U_{0}$, 
\begin{eqnarray}
S(E) = - e^{ i(\phi^{opt}(E) + \theta_{r}) } \frac{ 1 -   \sqrt{R} e^{-i\phi^{opt}(E)} }{ 1 - \sqrt{R} e^{i\phi^{opt}(E)} } \,,
\label{16-02} 
\end{eqnarray} 
and the factor 
\begin{eqnarray}
A_{n} = \frac{ \bar T  S(E)  e^{ i \pi \frac{n \hbar\Omega  }{\Delta }  }\,\frac{\sin\left( \pi \frac{n \hbar\Omega }{\Delta } \right) }{\pi n   }  }{\left( 1 - \sqrt{R} e^{-i\phi^{opt}(E)}  \right) \left( 1 - \sqrt{R} e^{i(\phi^{opt}(E) + n\Omega\tau  )}  \right) }  .
\nonumber \\
\label{16-01} 
\end{eqnarray} 

For practical calculations it is convenient to exploit the fact that under the   conditions set out in Eq.~(\ref{08}) the energy quantum $\hbar\Omega$ is much smaller than any other energy scale over which the Floquet scattering matrix changes.
Then one can ignore the discreteness on $n$ and introduce a continuous variable $\Omega_{n} = n\Omega$ instead.

\subsubsection{Continuous frequency representation}
\label{cfr}

The following relations establish correspondence between the discrete frequency representation we used before and the continuous frequency representation we are going to use from now on:  

\begin{eqnarray}
\sum\limits_{n = - \infty}^{\infty} &\to& \int\limits_{-\infty}^{\infty} \frac{d\Omega_{n} }{\Omega } \,,  
\nonumber \\
  \int\limits_{0}^{\cal T} dt^{\prime}  e^{in\Omega t^{\prime}} &\to& \int\limits_{-\infty}^{\infty} dt^{\prime}  e^{i\Omega_{n} t^{\prime}  } \,. 
\label{17} 
\end{eqnarray}
\ \\
\noindent
In addition the Kronecker symbol has to be replaced with the Dirac delta function, $\delta_{n,m} \to \Omega \delta(\Omega_{n} - \Omega_{m})$. 

To simplify notations even more we normalize energies by the level spacing, 

\begin{equation}
\epsilon = \frac{E-\mu  }{ \Delta }\,, \hspace{1.5cm} \omega_{n} = \frac{\hbar\Omega_{n}  }{ \Delta } \,.
\label{18}
\end{equation}
\ \\
\noindent
Then, for any function $\psi(E)$ after the substitutions given in Eq.~(\ref{18}) we use the notation $\psi(\epsilon)$. 
For instance, the Fermi function $f(E) = \left( 1 + e^{ \frac{ E - \mu }{ k_{B} T}}  \right)^{-1}$ in new variables reads as follows,    

\begin{eqnarray}
f(\epsilon) = \frac{1 }{  1 + e^{ \frac{ \epsilon }{ k_{B} T/\Delta} }  }  \,,
\label{contfermi}
\end{eqnarray}
\ \\
\noindent
where $k_{B}$ is the Boltzmann constant.  
In addition we use the following short notation,

\begin{equation}
\rho(\epsilon) = \frac{ 1 - \sqrt{R} e^{i \phi^{opt}(\epsilon)} }{\sqrt{\bar T} } \,, 
\label{19}
\end{equation}
\ \\
\noindent
with $\phi^{opt}(\epsilon) = \pi + 2\pi \epsilon$, see Eq.~(\ref{14}). 
With new variables and short notations the equation (\ref{16}) becomes

\begin{eqnarray}
S_{F}^{opt}\left(\epsilon, \epsilon + \omega_{n} \right) = S(\epsilon) \frac{ \hbar\Omega }{\Delta }  \frac{  \sin(\pi  \omega_{n})  }{\pi \omega_{n}  }\, e^{ i\pi \omega_{n} }  
\nonumber \\
\times \left\{ \delta(\omega_{n})   -    \frac{ \frac{ \exp \left( -i 2\pi \omega_{n}\frac{ t_{-} }{\tau }   \right)  }{\left( 1 + \omega_{n} \right) } +  \frac{  \exp \left( -i 2\pi \omega_{n}\frac{ t_{+} }{\tau }   \right)  }{\left( 1 - \omega_{n} \right) } }{ \rho^{*}(\epsilon) \rho(\epsilon + \omega_{n}) }  \right\}\,. 
\label{20} 
\end{eqnarray}
\ \\
\noindent
We remind that $\Omega$ is the frequency of the drive,  $\Delta$ is the level spacing, $t_{-}$ ($t_{+}$) is the time of an electron (hole) emission.
In Appendix \ref{uni} we show that this Floquet scattering matrix is unitary.  

Using above equation in Eq.~(\ref{i2t}) we calculate a time dependent current which consists of asymmetric pulses decaying exponentially, \cite{Feve:2007jx,Moskalets:2008fz,Keeling:2008ft} (see Fig.~\ref{fig3})

\begin{eqnarray}
I^{na}(t) = 
\frac{{\rm e}\theta(t - t_{-}) }{\tau_{D} } \, e^{-\frac{t - t_{-} }{\tau_{D} }} 
-
\frac{{\rm e} \theta(t - t_{+}) }{\tau_{D} } \, e^{-\frac{t - t_{+} }{\tau_{D} }} 
\,.
\label{itnad}
\end{eqnarray}
\ \\
\noindent
Here the superscript ``$na$'' stands for the non-adiabatic regime; $\theta(t)$ is the Heaviside step function

\section{Excess noise}
\label{en}

\subsection{Adiabatic regime}
\label{enar}

The adiabatic regime implies that the Floquet scattering matrix changes only a little with energy on the scale put by the frequency, i.e., on the scale $\hbar\Omega \sim \hbar\omega$. \cite{Moskalets:2002hu} 
Therefore, we can expand the scattering amplitudes in the powers of $\omega$. 
Since the capacitor does not support a DC current, the current correlations function tends to zero at zero frequency. \cite{Buttiker:1993wh} 
Taking into account the unitarity of the Floquet scattering matrix, Eq.~(\ref{21}), we find from Eq.~(\ref{33}) that the zero-frequency noise, $\omega=0$, is indeed zero. 

At small frequency we expand $\Pi_{m}(E,\omega)$, Eq.~(\ref{34}), up to the second order in $\omega$  and find,

\begin{eqnarray}
\left |\Pi_{m}\left( E,\omega \right) \right|^{2} =  
\delta_{m,0} \left( 1 + \omega^{2}  {\rm Re} \frac{\partial^{2} \Pi_{m}\left( E,\omega \right) }{\partial \omega^{2} }  \Big |_{\omega = 0}
 \right)
\nonumber \\
+ \omega^{2} \left | \frac{\partial \Pi_{m}\left( E,\omega \right) }{\partial \omega }  \Big |_{\omega = 0}\right |^{2}
+ {\cal O}(\omega^{3})
\,. 
\label{35} 
\end{eqnarray}
\ \\
\noindent
Here ${\cal O}(\omega^{3})$ denotes the third and higher order in $\omega$ terms. 
Note, due to the unitarity condition, Eq.~(\ref{21}), the linear in $\omega$ term vanishes in the above equation.
As it follows quite generally from Eqs.~(\ref{33}) and (\ref{35}), the low frequency noise is quadratic in $\omega$.
Earlier the similar behavior was  shown for the noise of a chaotic cavity. \cite{Brouwer:1997fo}

To the leading order in $\Omega$ the Floquet scattering matrix elements are given by the Fourier coefficients of the frozen scattering amplitude. \cite{Moskalets:2011cw}
Thus we use 

\begin{eqnarray}
S_{F}\left(E_{q}, E\right) \approx S_{q}(E) \,,
\label{adfs}
\end{eqnarray}
\ \\
\noindent
in Eqs.~(\ref{35}), (\ref{34}) and then in Eq.~(\ref{33}) and find,

\begin{eqnarray}
{\cal P}^{ex,ad}(\omega) = \frac{{\rm e}^{2} \hbar\omega^{2} }{2\pi} \int\limits_0^\infty  dE  \sum\limits_{m =  - \infty }^\infty  \left |\left ( S \frac{\partial S^{*} }{\partial E } \right )_{m}  \right |^{2} 
 \nonumber \\
\times \left\{    F ( {E ,E + m \hbar\Omega  + \hbar \omega } )   -   F ( {E ,E  + \hbar \omega } )   \right\} .
\label{36} 
\end{eqnarray} 
\ \\
\noindent
where the subscript ``$m$'' indicates the Fourier coefficient. 
Note, while calculating the above equation we used the following identities,

\begin{eqnarray}
{\rm Re}\, S \frac{\partial^{2} S^{*} }{\partial E^{2} } &=& - \frac{\partial S }{ \partial E} \frac{ \partial S^{*} }{ \partial E } ,
\nonumber \\
\sum\limits_{m}  \left | \left ( S \frac{\partial S^{*} }{\partial E } \right )_{m} \right |^{2}  &=&  \left ( \frac{\partial S }{\partial E } \frac{\partial S^{*} }{\partial E }  \right )_{0} .
\label{37} 
\end{eqnarray}
\ \\
\noindent
which can be proven using the unitarity, $|S(E,t)|^{2} = 1$. 

If the temperature is of the order of the frequency, $k_{B} T \sim \hbar\omega \sim \hbar\Omega$, then the scattering amplitude can be kept constant over the entire energy interval relevant for the integration in Eq.~(\ref{36}). 
This follows from the fact mentioned already that in the adiabatic regime the scattering matrix can be kept as energy-independent on the scale of order $\hbar\Omega$.  
Therefore, we can easily integrate in Eq.~(\ref{36}): 

\begin{eqnarray}
{\cal P}^{ex,ad}(\omega) = \pi {\rm e}^{2} \hbar^{2} \omega^{2}    \sum\limits_{m =  - \infty }^\infty |\nu_{m}|^{2}  
\nonumber \\
\times \left\{ \! \left( m\Omega + \omega \right)  \coth\left( \frac{m\Omega + \omega }{2k_{B} T/\hbar } \right)   - \omega   \coth\left( \frac{\omega }{2k_{B} T/\hbar } \right)\!   \right\} .
 \nonumber \\
\label{38} 
\end{eqnarray} 
\ \\
\noindent
Here $\nu_{m}$ is the $m$th Fourier coefficient of the frozen density of states of electrons in the cavity, 

\begin{eqnarray}
\nu(E,t) = \frac{i }{2\pi} S(E,t) \frac{\partial S^{*}(E,t)  }{ \partial E}\,,
\label{dos}
\end{eqnarray}
\ \\
\noindent
evaluated at the Fermi energy, $E = \mu$. 

In the quantized emission regime, if the amplitude of the potential $U(t)$ is chosen such that only one level crosses the Fermi energy [e.g., $n=0$ in Eq.~(\ref{06-1})], we find the Fourier coefficients of the frozen scattering matrix

\begin{equation}
S_{q} = -\, 2\Omega \Gamma_{\tau}\, e^{-|q|\Omega \Gamma_{\tau}}\, e^{i\theta_{r}} \left\{
\begin{array}{ll}
e^{i q \Omega t_{-} } \,, & q > 0\,, \\
\ \\
e^{iq\Omega t_{+} } \,, & q < 0\,.
\label{07_nsc_10}
\end{array}
\right.
\end{equation}
\ \\
\noindent
and calculate the excess noise power, \cite{Moskalets:2013tl}
  
\begin{eqnarray}
{\cal P}^{ex,ad}(\omega) = \frac{2 {\rm e}^{2} }{\pi } \left( \omega \tau_{D} \right)^{2} \left( \Omega \Gamma_{\tau} \right)^{2}   \sum\limits_{m =  - \infty }^\infty  e^{-2|m|\Omega\Gamma_{\tau}}  
\nonumber \\
\times \left\{\!  \left( m\Omega + \omega \right)  \coth\left( \frac{m\Omega + \omega }{2k_{B} T/\hbar } \right)   - \omega   \coth\left( \frac{\omega }{2k_{B} T/\hbar } \right) \!  \right\} ,
 \nonumber \\
\label{39} 
\end{eqnarray} 
\ \\
\noindent
where the dwell time $\tau_{D} = h/(\bar T \Delta)$. 
Above equation is even in frequency, ${\cal P}^{ex,ad}(\omega) = {\cal P}^{ex,ad}(-\omega)$, as it should be accordingly to Eq.~(\ref{45-1}) taken at $\ell=0$. 

Note, for the case when the Fermi level aligns  with a quantum level in the cavity at $U=0$ and the amplitude of the time-dependent potential is equal to the level spacing, ${\rm e}U(t) = (\Delta/2) \cos\left( \Omega t \right)$,  the half-duration of an emitted current pulse $\Gamma_{\tau} \sim \bar\delta/\left| {\rm e}dU/dt \right|$ becomes  

\begin{eqnarray}
\frac{ \Gamma_{\tau}  }{\cal T } = \frac{ \bar T }{ 4\pi^{2} } 
\,.
\label{gamma}
\end{eqnarray}
\ \\
\noindent 
Remember ${\cal T} = 2 \pi/\Omega$ is the period of the drive and $\bar T\ll 1$ is the transmission probability of the QPC connecting the cavity and the electron waveguide.   
Note the adiabaticity condition, Eq.~(\ref{adcond}), now reads, $\bar T^{2} \gg 4\pi^{2} \hbar \Omega/\Delta$. \cite{Splettstoesser:2008gc}

\subsubsection{Low temperatures}
\label{lt}

At low temperatures, $k_{B} T \ll \hbar\omega, \hbar\Omega$, the hyperbolic cotangent  in  Eq.~(\ref{39}) equals to $1$ with an exponential accuracy and  we can sum up over $m$,

\begin{eqnarray}
{\cal P}^{ex,ad}(\omega) &=& 2\frac{{\rm e}^{2} }{\cal T } \left( \omega\tau_{D} \right)^{2} e^{- 2 |\omega| \Gamma_{\tau}} .
\label{42} 
\end{eqnarray}
\ \\
\noindent
The factor $2$ in front is the number of particles (one electro and one hole) emitted during the period ${\cal T} = 2\pi /\Omega$. 
The electrons and holes are emitted at different times and, therefore, they  contribute to noise independently. 

\begin{figure}[t]
\begin{center}
\includegraphics[width=80mm]{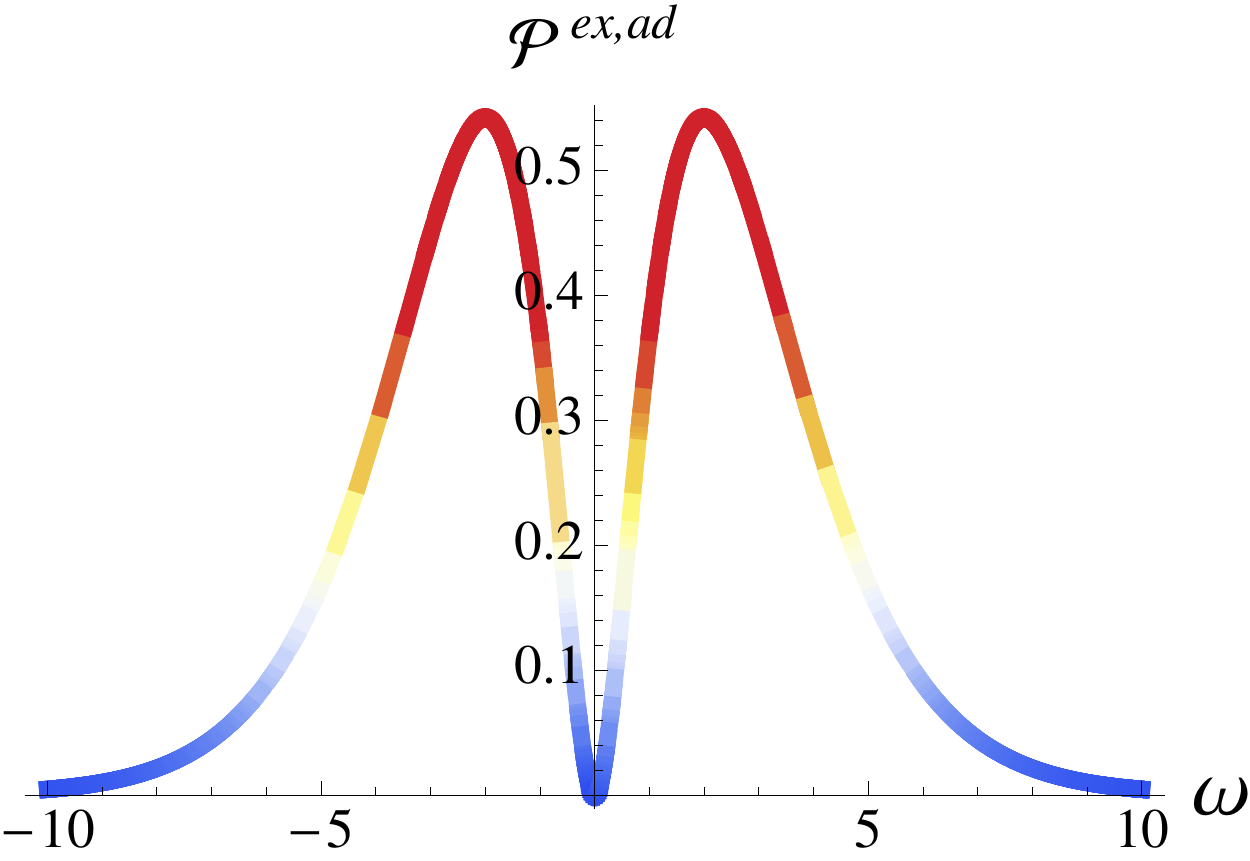}
\caption{(Color online)
Excess noise power in the adiabatic regime, Eq.~(\ref{42}), in units of  $2({\rm e}^{2}/{\cal T }) \left[ \tau_{D} /(2\Gamma_{\tau} ) \right]^{2}$. The frequency $\omega$ is in units of $1/(2\Gamma_{\tau})$. $\tau_{D}$ is the dwell time.  $\Gamma_{\tau}$ is the half-width of an emitted current pulse. The mean energy of emitted particles ${\cal E} = \hbar/(2\Gamma_{\tau})$ defines the frequency cut-off over which the excess noise power drops quickly.  The temperature is zero.
}
\label{fig4}
\end{center}
\end{figure}

The frequency dependence of the excess noise in the adiabatic regime is shown in Fig.~\ref{fig4}. 
At small frequencies, $\omega\Gamma_{\tau} \ll 1$, the noise is quadratic in frequency,

\begin{eqnarray}
{\cal P}^{ex,ad}_{0}(\omega) = 2\frac{{\rm e}^{2} }{\cal T } \left( \omega\tau_{D} \right)^{2}. 
\label{43-0}
\end{eqnarray}
\ \\
\noindent
At high frequencies, $\omega\Gamma_{\tau} \gg 1$, it gets suppressed.

\subsubsection{High frequency cut-off}
\label{hfco}

To explain the noise suppression effect we note the following.
Accordingly to the numerical calculations of  Ref.~\onlinecite{Parmentier:2012ed} carried out  for the  non-adiabatic regime, the single-particle excess noise is cut off for frequencies exceeding the energy of emitted particle (counted above the Fermi energy for emitted electrons and below the Fermi energy for emitted holes). 
For simplicity we consider the zero temperature case only. 

To account for this effect we calculate how many electrons (on average) have an energy larger than the energy quantum $\hbar\omega$ and, therefore, are able to emit it. 
We denote this number as ${\cal N}_{e}^{ad}(\omega)$.  
To calculate it we need the probability $p_{e}^{ad}(E)$ for a single electron to be emitted with energy $E$. 
These two quantities are related as follows,   

\begin{eqnarray}
{\cal N}_{e}^{ad}(\omega) = \int\limits_{\mu + \hbar\omega}^{\infty} dE p^{ad}_{e}(E) \,.
\label{np}
\end{eqnarray}
\ \\
\noindent
Since in the quantized emission regime only a single particle can be emitted at a time,   the probability $p_{e}^{ad}(E)$ is directly related to the distribution function $f_{e}^{ad}(E)$ of electrons scattered off the dynamic cavity, see Sec.~\ref{dis-fun}: 

\begin{eqnarray}
p_{e}^{ad}(E) = \frac{f_{e}^{ad}(E) }{ \int\limits_{0}^{\infty} dE f_{e}^{ad}(E) } \,.
\label{pf}
\end{eqnarray} 
\ \\
\noindent
Using the distribution function given in Eq.~(\ref{a01}), we calculate the probability,

\begin{eqnarray}
p_{e}^{ad}(E) = \frac{2 \Gamma_{\tau}  }{ \hbar }  \, e^{ - \frac{2\Gamma_{\tau} }{\hbar }\, |E - \mu| }\,.
\label{pad}
\end{eqnarray}
\ \\
\noindent
Using this probability we can evaluate the mean energy (counted from the Fermi energy) of emitted electrons, ${\cal E} = \hbar/(2\Gamma_{\tau})$, which was found in Ref.~\onlinecite{Moskalets:2009dk} on the base of the dc heat carried by the emitted electrons.

Using Eq.~(\ref{pad}) in Eq.~(\ref{np}) we find 
\begin{eqnarray}
{\cal N}_{e}^{ad}(\omega) = e^{ - 2\Gamma_{\tau} |\omega | } \,.
\label{Nad}
\end{eqnarray}
\ \\
\noindent
For the holes we get the same answer, ${\cal N}_{h}^{ad}(\omega) = {\cal N}_{e}^{ad}(\omega)$. 

Thus, the zero temperature result (\ref{42}) can be represented as the phase noise of two emitted particles ${\cal P}^{ex,ad}_{0}(\omega)$, Eq.~(\ref{43-0}), times by the  efficiency  factor ${\cal N}^{ad}(\omega) \equiv {\cal N}_{e}^{ad}(\omega) = {\cal N}_{h}^{ad}(\omega)$.

\subsection{Non-adiabatic regime}
\label{enpsp}

If the cavity is driven by the pulsed potential, Eq.~(\ref{07}), the excess noise persists up to high frequencies and we cannot use the expansion given in Eq.~(\ref{35}). 
Instead, we use directly Eqs.~(\ref{33}) and (\ref{34}).
In the continuous frequency representation, see Sec.~\ref{cfr}, the excess noise power reads, 

\begin{eqnarray}
{\cal P}^{ex}(\bar\omega) &=& \frac{{\rm e}^{2} \Omega }{2\pi} \int\limits_{-\infty}^\infty  d\epsilon \int\limits_{-\infty}^\infty d\omega_{m}    F ( {\epsilon ,\epsilon + \omega_{m}  + \bar\omega } )  
\nonumber \\
&& \times \left\{ \left | \Pi_{m}(\epsilon,\bar\omega) \right |^{2}  -  \delta^{2}\left( \omega_{m} \right)  \right\} ,
 \nonumber \\
\Pi_{m}(\epsilon,\bar\omega) &=&  \left( \frac{\Delta }{\hbar\Omega } \right)^{2} \int\limits_{-\infty}^\infty d\omega_{q}\,  S_{F} \left( \epsilon + \omega_{q},\epsilon \right) 
 \nonumber \\
&&\times S_{F}^* \left( \epsilon + \bar\omega + \omega_{q}  ,\epsilon + \bar\omega + \omega_{m}  \right) .
\label{39-1} 
\end{eqnarray}
\ \\
\noindent
By analogy with Eq.~(\ref{18}) here we introduced the  dimensionless frequency  $\bar\omega =  \hbar\omega/\Delta$. 

Note in the quantized emission regime under optimal working conditions the excess noise reaches its minimal value and is referred to as {\it the phase noise}. 
As we already mention the phase noise is due to the uncertainty in the single-particle emission time.\cite{Mahe:2010cp,Albert:2010co,Parmentier:2012ed} 

\subsubsection{Finite temperatures}
\label{fftt}

We calculate the excess noise power of the cavity working under optimal operating conditions, see Sec.~\ref{optimal}, and weakly coupled to an electron waveguide, $\bar T \ll 1$. 
We use the scattering matrix $S_{F}^{opt}$, Eq.~(\ref{20}), and obtain after lengthy but straightforward calculations (see Appendix \ref{app1}),   

\begin{eqnarray}
{\cal P}^{ex,na}(\omega)   = 2\frac{{\rm e}^{2} }{\cal T }  
\frac{\omega^{2}\tau_{D}^{2} }{1 + \omega^{2}\tau_{D}^{2} } {\cal N}^{na}_{T}(\omega) ,
\label{74}
\end{eqnarray}
\ \\
\noindent
with ${\cal T} = 2\pi/\Omega$ the period of the drive, $\tau_{D}$ the dwell time, and the finite-temperature (the subscript ``$T$'') efficiency factor 

\begin{eqnarray}
{\cal N}^{na}_{T}(\bar\omega) = 
\sum\limits_{a=-\infty}^{\infty}
\big\{ 
- 2F ( a - 0.5, a - 0.5 + \bar\omega  ) 
\nonumber \\
+ F ( a - 0.5, a + 0.5 + \bar\omega  ) + F ( a - 0.5, a + 0.5 - \bar\omega  ) 
\big\}
\,.
\nonumber \\
\label{75} 
\end{eqnarray}
\ \\
\noindent
Notice the arguments of the function $F$, Eq.~(\ref{002}), in above equation are normalized accordingly to Eq.~(\ref{18}). 
Using Eq.~(\ref{fsym}) one can check that ${\cal N}^{na}_{T}(\bar\omega) = {\cal N}^{na}_{T}(-\bar\omega)$ and hence the excess noise power is even in frequency, ${\cal P}^{ex,na}(\omega)  = {\cal P}^{ex,na}(-\omega) $ in agreement with Eq.~(\ref{45-1}) for $\ell=0$.  
  
The factor $2/{\cal T}$ in front of Eq.~(\ref{74}) accounts for the particle rate emission. 
The factor ${\rm e}^{2} \omega^{2}\tau_{D}^{2}/ \left( 1 + \omega^{2}\tau_{D}^{2} \right)$ accounts for the noise of a single particle. 
And finally the efficiency factor ${\cal N}^{na}_{T}(\omega)$ accounts for the fraction of particles able to emit the energy quantum $\hbar\omega$. 
Equation (\ref{75}) was derived for the vanishing level width, $\bar\delta \ll \Delta, k_{B}T$.  

\begin{figure}[b]
\begin{center}
\includegraphics[width=80mm]{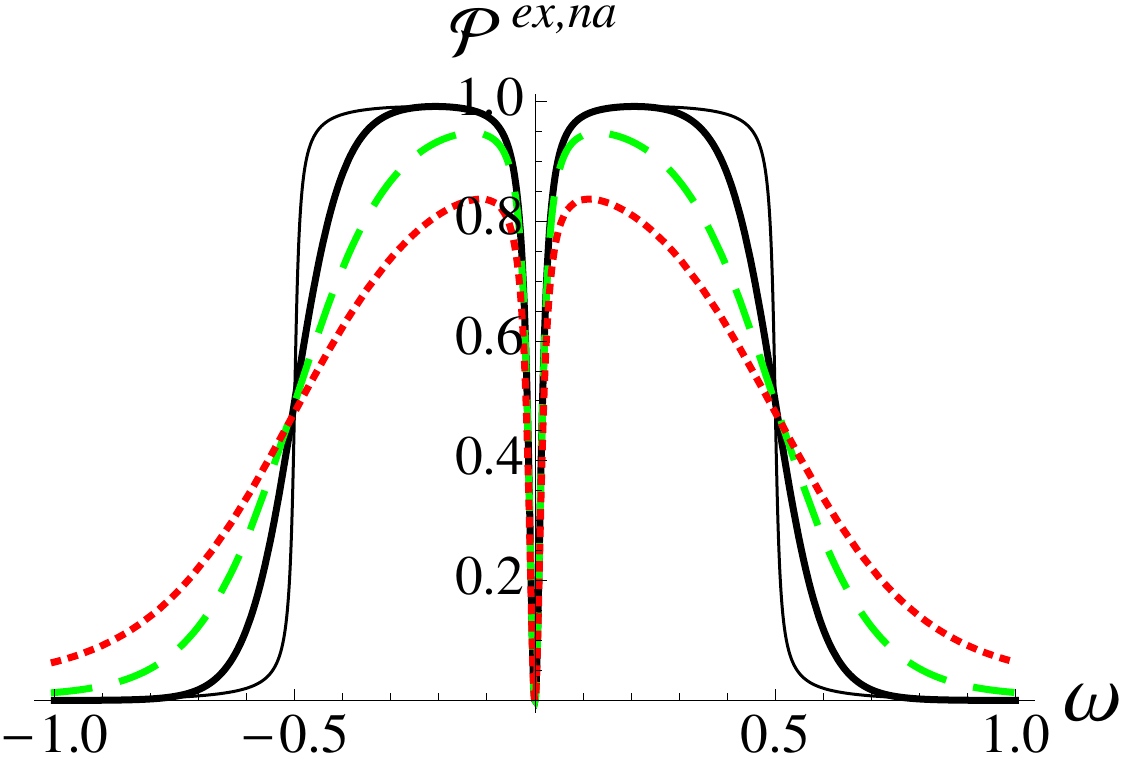}
\caption{(Color online)
Excess noise power in the non-adiabatic regime, Eq.~(\ref{74}), in units of $ 2 ({\rm e}^{2} /{\cal T })$. The frequency $\omega$ is in units of $\Delta/\hbar$. The noise power is given for different temperatures: $k_{B}T/\Delta = 0.05$ (black solid line), $0.1$ (green dashed line), and $0.15$ (red dotted line). The thin black solid line is the noise power  calculated with a zero-temperature efficiency factor, Eq.~(\ref{Nad-1}).  The cut-off frequency at $\omega \sim 1/2$ is due to the energy $\sim \Delta/2$ of emitted particles. The transmission of the cavity's QPC $\bar T = 0.1$ and, correspondingly, the level width is $\bar\delta = 0.008\Delta$. The dwell time $\tau_{D}\Delta/\hbar \approx 62.8$. 
}
\label{fig5}
\end{center}
\end{figure}

The frequency dependence of the excess noise power in the non-adiabatic regime is show in Fig.~\ref{fig5} for several temperatures. 
This agrees qualitatively with the Floquet numerical calculations present in  Ref.~\onlinecite{Parmentier:2012ed}. 
At low temperatures the noise is cut at $\omega > \Delta/(2\hbar)$. 
As in the adiabatic case the frequency cut-off is related to the energy of emitted particles, $\sim \Delta/2$ (counted from the Fermi energy $\mu$). 
Since all the states below $\mu$ are fully occupied, an electron cannot emit more energy than $\Delta/2$. 
However, with increasing temperature the states below the Fermi level become partially unoccupied. 
As a consequence, after emission an electron can jump below $\mu$ and, therefore, it can emit more energy than at zero temperature. 
This explains why with increasing temperature the noise persists up to higher frequencies and why the decay of the noise power at high frequencies is governed by the temperature.

\subsubsection{High-frequency cut-off at zero temperature}
\label{hfcozt}

The efficiency factor ${\cal N}^{na}_{T}(\omega)$, Eq.~(\ref{75}), was calculated in the limit of vanishing width of the quantum levels in the cavity, $\bar\delta \to 0$. 
In fact this limit is a high temperature limit, $k_{B}T \gg \bar\delta$. 
To estimate the efficiency factor at zero temperature we follow the same procedure as in the adiabatic regime, see Sec.~\ref{hfco} and calculate the fraction of emitted particles ${\cal N}_{e}^{na}(\omega)$ which are able to emit the energy quantum $\hbar\omega$. 
For this purpose we use the distribution function for the non-adiabatic regime, Eq.~(\ref{dfe02-2}), and find the corresponding probability,

\begin{eqnarray}
p_{e}^{na}(E) =  C \frac{\bar\delta/\pi }{ (E - \mu - \Delta/2)^{2} + \bar\delta^{2} } \,, \ \  0 < E-\mu < \Delta \,.
\label{dfapp}
\end{eqnarray}
\ \\
\noindent
The factor $C = \pi/(2  \arctan[\Delta/(2\bar\delta)])$ accounts for the normalization. 
Then we calculate, 

\begin{eqnarray}
{\cal N}_{e}^{na}(\omega) \equiv \!\!\! \int\limits_{\mu + \hbar\omega}^{\mu + \Delta} \!\!\! \!\! dE p_{e}^{na}(E) = \frac{1 }{2 } - \frac{ \arctan\left(2\omega\tau_{D} - \frac{\Delta }{ 2\bar\delta} \right) }{2 \arctan\left(\frac{\Delta }{ 2\bar\delta} \right)  } .
\label{Nad-1}
\end{eqnarray}
\ \\
\noindent
For the emitted holes the efficiency factor is the same.  
The frequency dependence of the noise power, Eq.~(\ref{74}) with ${\cal N}_{T}^{na}$ replaced by ${\cal N}_{e}^{na}$, is shown in Fig.~\ref{fig5} by the thin black solid line.


\section{Photon-induced noise}
\label{pin}

\subsection{Adiabatic regime}

Unlike the case considered in Sec.~\ref{enar}, the leading order term of the correlation function for $\ell \ne 0$ is linear in $\omega$. 
Therefore, to calculate the noise power, which is bi-quadratic in $\omega$ and $\Omega$, we have to keep the first order in $\Omega$ term in the expansion of the Floquet scattering matrix. 
Instead of Eq.~(\ref{adfs}) we now use the following, \cite{Moskalets:2009dk} 

\begin{eqnarray}
S_{F}(E_{p+\ell},E) &\approx& S_{p+\ell}(E) + \frac{(p+\ell)\hbar\Omega }{2 } \frac{\partial S_{p+\ell}(E) }{\partial E } ,
\nonumber \\
\nonumber \\
S_{F}^{*}(E_{p}^{\prime},E_{m}^{\prime}) &\approx& (S^{*})_{m-p}(E) 
\nonumber \\
\nonumber \\
&&+ \frac{(p+m)\hbar\Omega + 2\hbar\omega }{2 } \frac{\partial (S^{*})_{m-p}(E) }{\partial E }  , \quad\quad
\label{46-1} 
\end{eqnarray}
\ \\
\noindent
where $E^{\prime} = E + \hbar\omega$ and $S_{n}(E)$ is the Fourier coefficient of the frozen scattering amplitude $S(E,t)$. 
Substituting above equations into Eq.~(\ref{n34}) we find,

\begin{eqnarray}
\Pi_{m}^{\ell}(\omega)  \approx    \Bigg\{ 1  - \frac{i\hbar }{2 } S \frac{\partial^{2} S^{*} }{\partial t \partial E }  + \frac{i\hbar }{2 } S^{*} \frac{\partial^{2} S }{\partial t \partial E } 
\nonumber \\
+ \left( \hbar\omega + m \hbar\Omega \right) S \frac{\partial S^{*} }{ \partial E }  \Bigg\}_{m+\ell} .
\label{47-1}  
\end{eqnarray}
\ \\
\noindent
Thus to the lowest order in frequency the PIN power, Eq.~(\ref{43}), reads,

\begin{eqnarray}
{\cal P}_{\ell }^{PIN,ad}(\omega)  = \frac{{\rm e}^{2}  }{2\pi}  \int\limits_0^\infty  dE \left\{   S^{*}\frac{\partial S }{ \partial E }  \right\}_{\ell}
\nonumber \\
\times  \left\{  F \left( {E ,E_{-\ell}  + \hbar \omega } \right) \omega +   F \left( {E ,E  + \hbar \omega } \right) \left( \ell \Omega - \omega  \right) \right\} .
\label{n48} 
\end{eqnarray}
\ \\
\noindent
Note in the equation above we used the identity

\begin{eqnarray}
\frac{i\hbar }{2 }\left\{ S^{*} \frac{\partial^{2} S }{\partial t \partial E } - S \frac{\partial^{2} S^{*} }{\partial t \partial E }   \right\}_{\ell} = \ell \hbar\Omega \left\{ S^{*} \frac{\partial S }{ \partial E }  \right\}_{\ell} ,
\label{49-1}
\end{eqnarray}
\ \\
\noindent
following from the unitarity of the frozen scattering amplitude, $|S(E,t)|^{2} = 1$. 

At low temperatures, $k_{B}T \ll \hbar\omega, \hbar\Omega$, we put $S(E,t) \approx S(\mu,t)$ and integrate over energy in Eq.~(\ref{n48}). 
In the regime when the capacitor works as a single-particle source we use the  scattering amplitude given in Eq.~(\ref{06-1}) (with one level of the cavity crossing the Fermi level) and calculate the PIN power as follows, 

\begin{eqnarray}
{\cal P}_{\ell }^{PIN,ad}(\omega)  &=& 
{\cal C}_{\ell}
\frac{{\rm e}^{2} }{\cal T } \Omega^{2} \Gamma_{\tau} \tau_{D}
e^{-|\ell|\Omega \Gamma_{\tau}} 
\xi_{\ell}\left( \frac{\omega }{\Omega } \right) 
\,,
\nonumber \\
\xi_{\ell}(x) &=&   x |x - \ell |  + \left(  \ell - x \right) |x|   
\, ,
\nonumber \\
{\cal C}_{\ell} &=& - i  \left\{ e^{i \ell \Omega t_{-}} + e^{i \ell \Omega t_{+}} \right\} 
\, .
\label{n50} 
\end{eqnarray}
\ \\
\noindent
Remember, $t_{-}$ ($t_{+}$) is the time of an electron (hole) emission, $\tau_{D}$ is the dwell time, and $\Gamma_{\tau}$ is the half-width of the current pulse corresponding to an emitted particle, an electron or a hole. 
If electrons and holes are emitted equally distributed in time, $t_{+} = t_{-} + {\cal T}/2$, then 

\begin{eqnarray}
{\cal C}_{\ell} &=& - i e^{i \ell \Omega t_{-}} \left\{ 1 + (-1)^{\ell} \right\} 
\,.
\label{cell}
\end{eqnarray}
\ \\
\noindent 
Thus the PIN power is zero for odd $\ell$. 
Under the same conditions the Fourier harmonics of a current,

\begin{eqnarray}
I_{\ell} = \int\limits_{0}^{\cal T} \frac{dt }{\cal T } e^{- i \ell \Omega t} I(t) \,,
\label{fit}
\end{eqnarray}
\ \\
\noindent 
are zero for  even $\ell$. 
For instance, for the adiabatic current $I^{ad}(t)$, Eq.~(\ref{itad}),  we have 

\begin{eqnarray}
I_{\ell}^{ad} = 
\frac{{\rm e}  }{\cal T }
 e^{-|\ell|\Omega \Gamma_{\tau}} e^{i \ell \Omega t_{-}} \left\{ 1 - (-1)^{\ell}   \right\} \,.
\label{fitad}
\end{eqnarray}
\ \\
\noindent
The even-odd alternation for the current Fourier harmonics demonstrates clearly the presence of two carriers having opposite charge, while the even-odd alternation for the PIN  power is a clear signature of statistical independence of emitted carriers and, therefore, is a signature of the quantized emission regime with no spurious electron-hole pairs emitted.

\begin{figure}[t]
\begin{center}
\includegraphics[width=80mm]{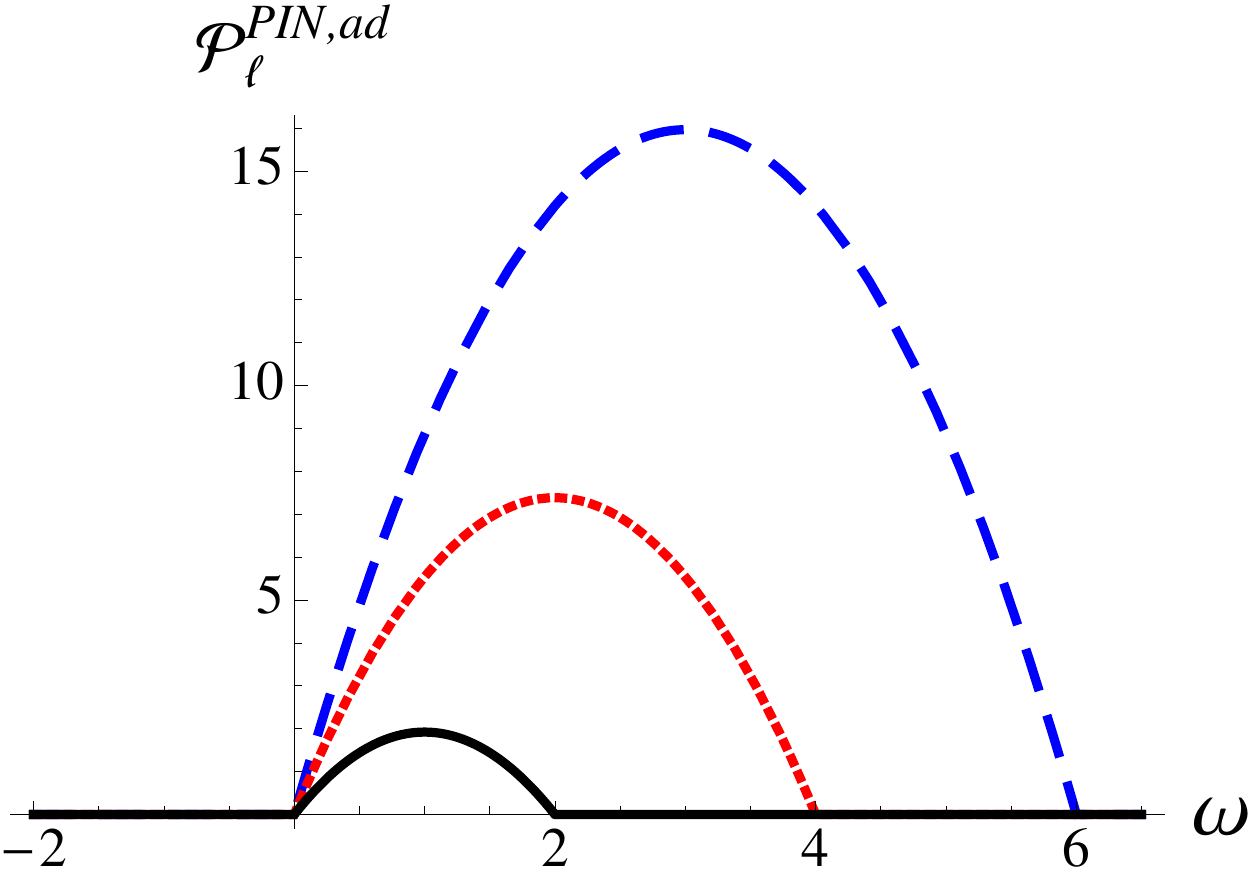}
\caption{
(Color online) 
Photon-induced noise power in the adiabatic regime.
The absolute value of the PIN power, Eqs.~(\ref{n50}) and (\ref{cell}), is shown in units of $2({\rm e}^{2}/{\cal T}) \Omega^{2} \tau_{D}\Gamma_{\tau}$  for different number of involved photons: $\ell = $ $2$ (black solid line), $4$ (read dotted line), and  $6$ (blue dashed line). 
The frequency $\omega$ is in units of the frequency of a drive $\Omega$.  
$\tau_{D}$ is the dwell time.  $\Gamma_{\tau}$ is the half-width of an emitted current pulse, which we choose here to be small such that $\Omega \Gamma_{\tau} \ll 1$. The temperature is zero. 
}
\label{fig6}
\end{center}
\end{figure}

The PIN power ${\cal P}_{\ell}^{PIN,ad}$, see Eqs.~(\ref{n50}) and (\ref{cell}), is shown in Fig.~\ref{fig6}. 
It differs essentially from the excess noise power ${\cal P}^{(ex,ad)}$, see  Eq.~(\ref{42}) and Fig.~\ref{fig4}. 
First, the PIN power is larger  by the factor of $\Gamma_{\tau}/\tau_{D} \sim (\Delta/\hbar\Omega) (\bar T/[2\pi])^{2} \gg 1$. 
Second, unlike the phase noise power, the PIN power is independent of the transparency $\bar T$ of the QPC connecting the cavity and the electron  waveguide provided that the adiabaticity condition, $\Gamma_{\tau} \gg \tau_{D}$, holds, see Eq.~(\ref{adcond}). 
It's maximum value is ${\cal P}_{\ell,max}^{PIN,ad} \sim ({\rm e}^{2} /{\cal T }) (\hbar\Omega/\Delta) (\ell^{2}/2)$. 
And, third, the PIN power is highly asymmetric in frequency: It exists within the finite frequency window, which is $0 < \omega < \ell \Omega$ for positive $\ell$, see Fig.~\ref{fig6}, and $0 > \omega > - |\ell | \Omega$ for negative $\ell$. 
Within its range of existence the PIN power is quadratic in frequency. 
The function $\xi_{\ell}\left( \frac{\omega }{\Omega } \right)$, defining the frequency dependent of the PIN power, reads,
\begin{equation}
\xi_{\ell}\left( \frac{\omega }{\Omega } \right)  = 
\left\{
\begin{array}{lll}
\dfrac{2 }{\Omega^{2} }
\omega \left( \ell\Omega - \omega \right) ; & \ell > 0 , & 0 < \omega < \ell \Omega , \\
\ & & \\
\dfrac{2 }{\Omega^{2} }
\omega \left( \omega - \ell\Omega  \right) ; & \ell < 0 , & - |\ell | \Omega < \omega < 0  . 
\end{array}
\right.
\label{panx05c} 
\end{equation}
\ \\
\noindent
Easy to check that Eq.~(\ref{n50}) satisfies the symmetries given in Eq.~(\ref{45-1}).


\subsection{Non-adiabatic regime}

For the potential $U(t)$, Eq.~(\ref{07}), driving the cavity non-adiabatically we use the continuous frequency representation, see Sec.~\ref{cfr}, and calculate the PIN power, Eq.~(\ref{43}), as follows, 

\begin{eqnarray}
{\cal P}_{\ell }(\bar\omega)  &=& \frac{{\rm e}^{2} \Omega }{2\pi} \int\limits_{-\infty}^\infty  d\epsilon  \int\limits_{ - \infty }^\infty d\omega_{m}   F \left( {\epsilon ,\epsilon + \omega_{m}  +  \bar\omega } \right) 
\nonumber \\
&& \times 
\Pi_{m}^{\ell}(\epsilon,\bar\omega)  \Pi_{m}^{0 *}(\epsilon,\bar\omega)
\, ,
\label{pan01cfr} 
 \end{eqnarray}
where
\begin{eqnarray}
\Pi_{m}^{\ell}(\epsilon,\bar\omega)  &=&  \left( \frac{\Delta }{\hbar\Omega } \right)^{2} \int\limits_{ - \infty }^\infty d \omega_{q}  S_{F}^{} \left( \epsilon + \omega_{q} + \omega_{\ell}  ,\epsilon  \right) 
\nonumber \\
&&\times S_{F }^* \left( \epsilon  + \bar\omega + \omega_{q},\epsilon  + \bar\omega + \omega_{m}   \right) 
\, ,
\label{pan03cfr} 
\end{eqnarray}
\ \\
\noindent 
where $\omega_{\ell} = \ell \hbar\Omega/\Delta$. 

To calculate above equations we use the Floquet scattering matrix $S_{F}^{opt}$, Eq.~(\ref{20}). 
Along the lines of Appendix \ref{app1} we calculate,

\begin{eqnarray}
{\cal P}_{\ell}^{PIN,na}(\omega)  = \frac{{\rm e}^{2} }{\cal T} e^{i \pi \frac{\ell\hbar\Omega }{\Delta }  } 
\frac{\sin\left( \pi \frac{\ell\hbar\Omega }{\Delta } \right) }{\pi \frac{\ell\hbar\Omega }{\Delta }\left( 1 - \left [ \frac{\ell\hbar\Omega }{\Delta } \right]^{2} \right) } 
\nonumber \\
\times
\frac{1 }{ \left( 1 - i \ell\Omega\tau_{D} \right)} 
\frac{ \omega \left( \omega -  \ell\Omega  \right) \tau_{D}^{2}     }{ \left( 1+  i \left[ \omega - \ell\Omega \right] \tau_{D} \right) \left( 1 - i \omega\tau_{D} \right)} 
\nonumber \\
\times
  \left\{ e^{ i \ell \Omega  t_{-}  }  {\cal N}_{\ell,-}(\omega) + e^{ i \ell \Omega  t_{+}  }  {\cal N}_{\ell,+}(\omega)  \right\} 
 \,,
\label{tpin} 
\end{eqnarray}
\ \\
\noindent    
where 

\begin{eqnarray}
{\cal N}_{\ell,\mp}(\bar\omega) = 
\sum\limits_{a = - \infty}^{\infty}
\Big\{
-
F(a - 0.5, a - 0.5 + \bar\omega  ) 
\nonumber \\
-
F(a - 0.5, a - 0.5 - \bar\omega + \omega_{\ell} ) 
 \nonumber \\
+
\frac{1 }{2 }
\left( 1 \mp \omega_{\ell} \right)
\big[ 
F(a - 0.5, a + 0.5 + \bar\omega )   
\nonumber \\
+ 
F( a - 0.5, a + 0.5 - \bar\omega + \omega_{\ell} )
 \big]
\nonumber \\
+
\frac{1 }{2 }
\left( 1 \pm \omega_{\ell} \right)
\big[ 
F(a - 0.5, a + 0.5 - \bar\omega )   
\nonumber \\
+
F( a - 0.5, a + 0.5 + \bar\omega - \omega_{\ell} )
\big ] 
\Big\}
\,.
\label{nellmp}
\end{eqnarray}
\ \\
\noindent
If electrons and holes are emitted at regular intervals, $t_{+} = t_{-} + {\cal T}/2$, then the PIN power, Eq.~(\ref{tpin}), is non-zero for even $\ell = 2\lambda$ only, while the current, Eq.~(\ref{itnad}), has only odd Fourier harmonics, 

\begin{eqnarray}
I_{\ell}^{na} = 
\frac{{\rm e}  }{\cal T }\, 
\frac{ e^{i \ell \Omega t_{-}} \left\{ 1 - (-1)^{\ell}   \right\} }{1 - i \ell \Omega \tau_{D} }  \,.
\label{fitnad}
\end{eqnarray}

\begin{figure}[t]
\begin{center}
\includegraphics[width=80mm]{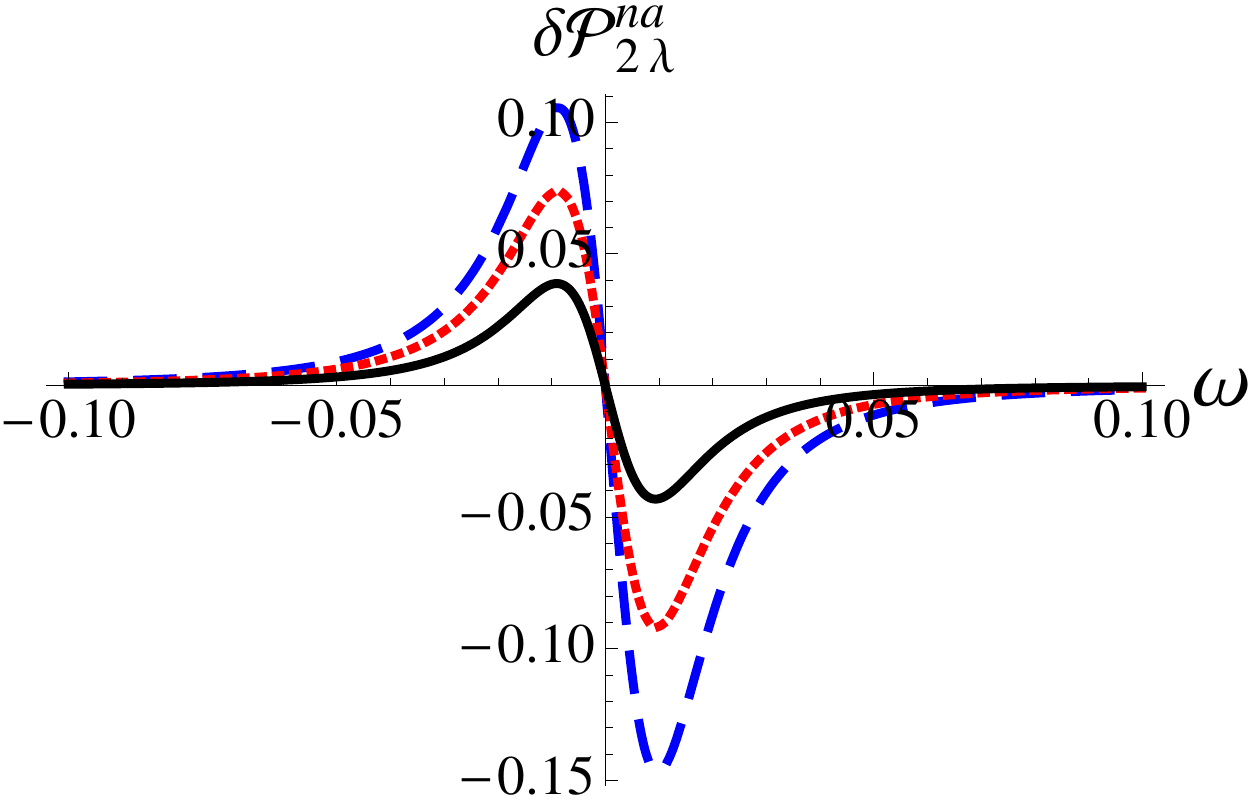}
\caption{
(Color online) 
Asymmetric part of the photon-induced noise power in the non-adiabatic regime, $\delta {\cal P}_{2\lambda}$, Eq.~(\ref{diff}), is plotted in units of $2({\rm e}^{2}/{\cal T})$  for different number of involved photons: $2\lambda = $ $2$ (black solid line), $4$ (read dotted line), and  $6$ (blue dashed line). 
The frequency $\omega$ is in units of $\Delta/\hbar$. 
The driving frequency $\Omega = 0.001 \Delta/\hbar$. 
The temperature is $k_{B}T/\Delta = 0.01$. 
Other parameters are the same as in Fig.~\ref{fig5}. 
 }
\label{fig7}
\end{center}
\end{figure}

The magnitude of ${\cal P}_{2\lambda}^{PIN,na}(\omega)$ is of the same order as the magnitude of the phase noise power ${\cal P}^{ex,na}(\omega)$, Eq.~(\ref{74}). 
This is easy to verify, since at $\ell = 0$ we have ${\cal N}_{\ell,-}(\bar\omega) = {\cal N}_{\ell,+}(\bar\omega) = {\cal N}_{T}^{na}(\bar\omega)$, see Eq.~(\ref{75}), and the equation (\ref{tpin}) becomes Eq.~(\ref{74}). 
However, in contrast to the phase noise power, the PIN power is asymmetric in frequency. 
To characterize this asymmetry we calculate the difference $\delta {\cal P}_{2\lambda}$ between the PIN power for some $2\lambda \ne 0$ and the phase noise power.  
For small $\ell$, such that $\ell \hbar\Omega/\Delta \ll 1$, we obtain the following: (for simplicity the origin of time is taken at the time of an electron emission, $t_{-} = 0$)

\begin{eqnarray}
&&\delta {\cal P}_{2\lambda}^{na}(\omega) \equiv {\cal P}_{2\lambda}^{PIN,na}(\omega) - {\cal P}^{ex,na}(\omega) \approx 
\nonumber \\
&&
\approx
2 \frac{{\rm e}^{2} }{\cal T} 
\left\{ 
\frac{ \omega \left( \omega -  2\lambda\Omega  \right) \tau_{D}^{2}      }{ 1+   \omega\left( \omega - 2\lambda\Omega \right) \tau_{D}^{2}  } 
-  
\frac{ \omega^{2} \tau_{D}^{2}      }{  1+   \omega^{2}\tau_{D}^{2}  } 
 \right\} 
\,.
\label{diff} 
\end{eqnarray}
\ \\
\noindent 
This function is plotted down in Fig.~\ref{fig7} for several positive $\lambda$. 
Its behavior for negative $\lambda$ can be anticipated on the base of the symmetry properties of the noise power, Eq.~(\ref{45-1}).

\section{Conclusion}
\label{concl}

Within the Floquet scattering matrix formalism I addressed  the finite-frequency noise spectrum of a quantum capacitor working as a single-particle emitter. 
I considered two working regimes, adiabatic and non-adiabatic. 
In the former/later case the potential driving the capacitor varies slow/fast on the scale of internal dynamics of the capacitor, which is characterized by the dwell time $\tau_{D}$, the time necessary for an electron to tunnel from the capacitor into the waveguide of vice versa. 
I calculated analytically the correlation function in the adiabatic regime and in the non-adiabatic regime under optimal operating conditions, see Sec.~\ref{optimal}. 

The current correlation function depends on two frequencies, $\omega_{1}$ and $\omega_{2}$. 
In the stationary case it is non-zero for $\omega_{1} = - \omega_{2}$ only.  
If the system (in our case the capacitor) is driven with frequency $\Omega$, then the correlation function becomes non-zero also for $\omega_{1} + \omega_{2} = \ell \Omega$, where $\ell$ is an integer. 
Therefore, the periodic drive manifests itself, first, by correlating currents at frequencies shifted by one or few frequencies of the drive and, second, by the modification of the correlation function for $\omega_{1} = - \omega_{2}$. 
The correlation function for $\ell \ne 0$ is referred to as the photon-induced noise power, since for the currents at $\omega_{1}$ and $\omega_{2} = - \omega_{1} + \ell\Omega$ to become correlated the electron system has to exchange $\ell$ energy quantum $\hbar\Omega$ with the external potential driving the capacitor. 
The difference between the correlation function for $\ell=0$ and the stationary one is referred to as the excess noise, since the drive generally increases the noise.  

The excess noise in the non-adiabatic regime was recently measured. \cite{Mahe:2010cp} 
In the quantized emission regime the excess noise reaches its minimal value (referred to as the phase noise), which is due to the quantum uncertainty in the emission time. \cite{Mahe:2010cp,Albert:2010co,Parmentier:2012ed}   
I calculated the excess noise power analytically in the non-adiabatic regime and found it to be suppressed at frequencies (times $\hbar$) exceeding the energy,  $\sim \Delta/2$,  of emitted particles, see Fig.~\ref{fig5}. 
This is in agreement with the results of numerical calculations of Ref.~\onlinecite{Parmentier:2012ed}. 
I predicted similar properties for the excess noise in the adiabatic regime with the trivial difference due to smaller characteristic energy, ${\cal E} = \hbar/(2\Gamma_{\tau}) \ll \Delta$, of emitted particles, see Fig.~\ref{fig4}.  

In contrast the photon-induced noise (PIN) shows striking difference for two discussed regimes. 
In the adiabatic regime the PIN power exists within the finite frequency window, see Fig.~\ref{fig6}, and its magnitude far exceeds the excess noise power, compare Eqs.~(\ref{n50}) and (\ref{42}). 
While in the non-adiabatic regime the PIN power deviates from the phase noise power only a little, compare Eqs.~(\ref{74}) and (\ref{tpin}). 
At the frequencies where it exists, the adiabatic PIN power is much larger than the non-adiabatic one. 
The last observation has a simple explanation. 
The PIN involves an energy exchange between the electron system and the driving time-periodic potential. 
This exchange is described by the Floquet scattering amplitude $S_{F}\left( E + \ell\hbar\Omega, E \right)$. 
In the adiabatic case $S_{F}^{ad} \sim \Gamma_{\tau}/{\cal T}$ provided that $\ell\Omega\Gamma_{\tau} \ll 1$, see Eq.~(\ref{07_nsc_10}). 
Here $\Gamma_{\tau}$ is the (half-)width of the current pulse corresponding to an adiabatically emitted electron and ${\cal T} = 2\pi/\Omega$ is the period of a drive.  
In the non-adiabatic case $S_{F}^{na} \sim \bar T \hbar\Omega/\Delta = \tau_{D}/{\cal T}$ provided that both $\ell\hbar\Omega/\Delta \ll 1$ and the level in the dot is not aligned with energy $E$, see Eq.~(\ref{16}). 
Remember that the dwell time $\tau_{D}$ defines also the width of the current pulse corresponding to an electron emitted non-adiabatically. 
The ratio of the scattering amplitudes, $S_{F}^{ad}/S_{F}^{na} \sim \Gamma_{\tau}/\tau_{D} \gg 1$, defines the ratio between the adiabatic PIN power, ${\cal P}_{\ell }^{PIN,ad} \sim \omega^{2}\tau_{D}\Gamma_{\tau}$, see Eq.~(\ref{n50}) and the non-adiabatic one, ${\cal P}_{\ell}^{PIN,na} \sim \omega^{2} \tau_{D}^{2}$, see  Eq.~(\ref{tpin}). 
Therefore, the PIN power provides a direct access to the Floquet scattering amplitudes. 

I also found that under optimal operating conditions the PIN power is zero for odd $\ell$ and is non-zero for even $\ell$. 
This parity-dependent behavior if being observed would indicate clearly the statistical independence of emitted electrons and holes and, therefore, would verify the quantized emission regime with no spurious electron-hole pairs emitted. 
Some advantage of the measurement of PIN compared to the measurement of the excess noise is that no subtracting procedures is needed, since the PIN is absent in the stationary case.

\begin{acknowledgments}
I thank Mathias Albert, Christian Flindt, and Markus B\"{u}ttiker for numerous helpful discussions. 
I appreciate the warm hospitality of the University of Geneva where part of this work was carried out. 
\end{acknowledgments}

\appendix

\section{Unitarity of the Floquet scattering matrix for the periodic step potential}
\label{uni}

I give the proof of unitarity since it is  helpful for subsequent calculations of the noise power. 
Here I follow close to Ref.~\onlinecite{Moskalets:2013dk}. 

The unitarity condition for a single orbital channel case reads, \cite{Moskalets:2004bo}

\begin{equation}
\sum\limits_{p = -\infty}^{\infty} S_{F}^{*}(E_{p}, E_{m}) S_{F}(E_{p}, E_{n}) = \delta_{m,n} \,,
\label{21} 
\end{equation}
\ \\
\noindent
where $p$, $m$, and $n$ all are integer. 
In the continuous frequency representation, see Sec.~\ref{cfr}, we rewrite above identity as follows,

\begin{eqnarray}
\int d\omega_{p} S_{F}^{*}(\epsilon + \omega_{p}, \epsilon + \omega_{m}) S_{F}(\epsilon + \omega_{p}, \epsilon + \omega_{n}) \nonumber \\
= \left( \frac{\hbar\Omega }{\Delta } \right)^{2} \delta(\omega_{m} - \omega_{n})\,. 
\label{22} 
\end{eqnarray}
\ \\
\noindent
Here we use dimensionless quantities accordingly to Eq.~(\ref{18}). 

For the scattering matrix given in Eq.~(\ref{20}) we find

\begin{eqnarray}
\int\limits_{-\infty}^{\infty}d\omega_{p} \frac{\sin(\pi [ \omega_{p} - \omega_{m}] ) \sin(\pi [ \omega_{p} - \omega_{n}])}{\pi^{2} [ \omega_{p} - \omega_{m}] [ \omega_{p} - \omega_{n}] }  e^{i \pi [\omega_{n} - \omega_{m}]} \nonumber \\
\times \left\{  \delta(\omega_{p} - \omega_{m}) - \frac{ \frac{ e^{-i 2\pi [\omega_{p} - \omega_{m}] \frac{t_{-} }{\tau } }}{1 + \omega_{m} - \omega_{p}  } + \frac{ e^{-i 2\pi [\omega_{p} - \omega_{m}] \frac{t_{+} }{\tau } }}{1 - \omega_{m} + \omega_{p} } }{\rho(\epsilon+\omega_{p}) \rho^{*}(\epsilon + \omega_{m}) }   \right\} \nonumber \\
\times \left\{  \delta(\omega_{p} - \omega_{n}) - \frac{ \frac{ e^{i 2\pi [\omega_{p} - \omega_{n}] \frac{t_{-} }{\tau } }}{1 +  \omega_{n} - \omega_{p}  } + \frac{ e^{i 2\pi [\omega_{p} - \omega_{n}] \frac{t_{+} }{\tau } }}{1 - \omega_{n} + \omega_{p} } }{\rho^{*}(\epsilon+\omega_{p}) \rho(\epsilon + \omega_{n}) }   \right\} \nonumber \\
= \delta(\omega_{m} - \omega_{n}) \,. \quad 
\label{23} 
\end{eqnarray}
\ \\
\noindent
Here we took into account that $|S|^{2} = 1$.
Next we open the curly brackets and integrate over $\omega_{p}$ using the Dirac delta function, 

\begin{eqnarray}
\int\limits_{-\infty}^{\infty} \frac{ d\omega_{p} }{|\rho(\epsilon+\omega_{p})|^{2} }  \frac{\sin(\pi [ \omega_{p} - \omega_{m}] ) \sin(\pi [ \omega_{p} - \omega_{n} ])}{\pi^{2} [ \omega_{p} - \omega_{m}] [ \omega_{p} - \omega_{n}] }  
\nonumber \\
\times \bigg\{ \xi_{p} +  \frac{e^{ -i  2\pi [ \omega_{n} - \omega_{m}] \frac{ t_{-} }{\tau }  }  }{ \left( \omega_{p} - \omega_{m}  -1 \right)\left( \omega_{p} - \omega_{n}  -1 \right) } \nonumber \\
 + \frac{e^{ -i 2\pi  [ \omega_{n} - \omega_{m}] \frac{ t_{+} }{\tau }  } }{ \left( \omega_{p} - \omega_{m}  +1 \right)\left( \omega_{p} - \omega_{n}  +1 \right) }  \bigg\} =  \frac{ 2 \sin(\pi  [\omega_{n} - \omega_{m}] )  }{\pi [\omega_{n} - \omega_{m}]  }  \nonumber \\
\nonumber \\
\times  \frac{  e^{ -i 2\pi  [ \omega_{n} - \omega_{m}] \frac{ t_{-} }{\tau }  } + e^{ -i 2\pi [ \omega_{n} - \omega_{m}] \frac{ t_{+} }{\tau }  }   }{1 -  [\omega_{n} - \omega_{m}]^{2}    }  
\,,
\quad
\label{24} 
\end{eqnarray}
with
\begin{eqnarray}
\xi_{p} = - e^{-i 2\pi \omega_{p} \frac{ t_{+} - t_{-} }{\tau } }  \frac{ e^{ - i  2\pi  \omega_{n} \frac{ t_{-} }{\tau }  }\, e^{ i  2\pi  \omega_{m} \frac{ t_{+} }{\tau }  }  }{ \left( \omega_{p} - \omega_{n} -1 \right)\left(\omega_{p} - \omega_{m}  +1 \right) } 
\nonumber \\
- e^{i 2\pi \omega_{p} \frac{ t_{+} - t_{-} }{\tau } }  \frac{ e^{ - i 2\pi  \omega_{n} \frac{ t_{+} }{\tau }  }\, e^{  i 2\pi  \omega_{m} \frac{ t_{-} }{\tau }  } }{ \left( \omega_{p} - \omega_{n}  +1 \right)\left( \omega_{p} - \omega_{m}  -1 \right) }  \,.
\label{25} 
\end{eqnarray}
\ \\
\noindent
Under the condition set out in Eq.~(\ref{08}) we see that $\xi_{p}$ oscillates fast with $\omega_{p}$. 
Therefore, it nullifies the corresponding integral. 
Physically it means that an electron emission at $t = t_{-}$ is completely independent of a hole emission taking place at $t = t_{+}$. 
In the rest of Eq.~(\ref{24}) one can separate the terms dependent on $t_{-}$ and $t_{+}$. 
Then what remains to show is that  

\begin{eqnarray}
\int\limits_{-\infty}^{\infty} \frac{ d\omega_{p} }{|\rho(\epsilon+\omega_{p})|^{2} }  \frac{\sin(\pi [ \omega_{p} - \omega_{m}] ) \sin(\pi [\omega_{p} - \omega_{n} ])}{\pi^{2} [ \omega_{p} - \omega_{m} ] [ \omega_{p} - \omega_{n} ] }  
\nonumber \\
\times   \frac{1  }{ \left( \omega_{p} - \omega_{m}  \mp 1 \right)\left( \omega_{p} - \omega_{n} \mp 1 \right) } =  \frac{ 2 \sin(\pi   \omega_{q} )  }{\pi \omega_{q} \left( 1 -  \omega_{q}^{2} \right) } \,,  
\label{26} 
\end{eqnarray}
\ \\
\noindent
where $\omega_{q} = \omega_{n} - \omega_{m}$. 
To prove above identity we utilize the periodicity of the function $\rho(\epsilon) = \rho(\epsilon + a)$, where $a$ is an integer, see Eq.~(\ref{19}), and represent the integral over $\omega_{p}$ as follows,

\begin{eqnarray}
\int\limits_{-\infty}^{\infty} d\omega_{p} = \sum\limits_{a = -\infty}^{\infty}\, \int\limits_{0}^{1} d\omega_{p}^{\prime} \,,
\label{27} 
\end{eqnarray}
\ \\
\noindent
with $\omega_{p} = \omega_{p}^{\prime} + a$. 
Then Eq.~(\ref{26}) becomes

\begin{eqnarray}
\int\limits_{0}^{1} d\omega_{p}^{\prime}\, \frac{ \Sigma_{q}  }{|\rho(\epsilon+\omega_{p}^{\prime})|^{2} }  =  \frac{ 2 \sin(\pi   \omega_{q} )  }{\pi \omega_{q} \left( 1 -  \omega_{q}^{2} \right) } \,,
\label{28} 
\end{eqnarray}
with 
\begin{eqnarray}
\Sigma_{q} =  \sum\limits_{a = -\infty}^{\infty}  \frac{\sin(\pi [ \omega_{p}^{\prime} - \omega_{m} ] ) \sin(\pi [ \omega_{p}^{\prime} - \omega_{n} ])}{\pi^{2} [ \omega_{p}^{\prime} - \omega_{m} + a] [ \omega_{p}^{\prime} - \omega_{n} + a] }  
\nonumber \\
\times   \frac{1  }{ \left( \omega_{p}^{\prime} - \omega_{m}  + a  \mp 1 \right)\left( \omega_{p}^{\prime} - \omega_{n} + a  \mp 1 \right) } 
\,.
\label{29} 
\end{eqnarray}
\ \\
\noindent
To calculate $\Sigma_{q}$ we use the following identity,

\begin{eqnarray}
 \sum\limits_{a=-\infty}^{\infty} \frac{1 }{\left\{(a  + y)^{2} - \frac{1}{4} \right\} \left\{ \left(a  + [x+y] \right)^{2} - \frac{1}{4} \right\}} 
\nonumber \\
=    \frac{\sin(\pi x)  }{ x \left( 1-  x^{2} \right)} \frac{2\pi }{\cos(\pi y) \cos(\pi[x+ y])  } \,, 
\label{30} 
\end{eqnarray}
\ \\
\noindent
with $y = \omega_{p}^{\prime} - \omega_{n} \mp 0.5$ and $x = \omega_{q}$. 
This identity can be proven by expanding the left hand side into the simple fractions and then using the following text-book sum  

\begin{eqnarray}
\sum\limits_{a=-\infty}^{\infty} \frac{1 }{a + \gamma } = \pi \cot(\pi \gamma)  \,, 
\label{31}
\end{eqnarray}
\ \\
\noindent
taken at proper $\gamma$'s. 
As a result we find

\begin{eqnarray}
\Sigma_{q} =  \frac{2\sin(\pi \omega_{q})  }{ \pi \omega_{q} \left( 1-  \omega_{q}^{2} \right)} \,.
\label{32}
\end{eqnarray}
\ \\
\noindent
Since $\Sigma_{q}$ is independent of $\omega_{p}^{\prime}$, we can integrate in Eq.~(\ref{28}). 
With $\rho$ given in Eq.~(\ref{19}) the integral is one. 
Thus the identity (\ref{28}) and hence the unitarity condition, Eq.~(\ref{22}), is proven.

Alternatively one can use the following unitarity condition, \cite{Moskalets:2004bo}

\begin{equation}
\sum\limits_{p = -\infty}^{\infty} S_{F}^{*}(E_{m}, E_{p}) S_{F}(E_{n}, E_{p}) = \delta_{m,n} \,,
\label{21-1} 
\end{equation}
\ \\
\noindent
which in the continuous frequency representation reads,

\begin{eqnarray}
\int d\omega_{p} S_{F}^{*}(\epsilon + \omega_{m}, \epsilon + \omega_{p}) S_{F}(\epsilon + \omega_{n}, \epsilon + \omega_{p}) \nonumber \\
= \left( \frac{\hbar\Omega }{\Delta } \right)^{2} \delta(\omega_{m} - \omega_{n})\,. 
\label{22-1} 
\end{eqnarray}
\ \\
\noindent
Here we use normalized quantities introduced in Eq.~(\ref{18}). 
We will use Eq.~(\ref{22-1}) below, see Sec.~\ref{nar}.

\section{Excess noise for the periodic step potential}
\label{app1}

We calculate the excess noise, Eq.~(\ref{39-1}), of the SES working under optimal operating conditions. 
The Floquet scattering matrix elements necessary for calculations are following, see Sec.~\ref{optimal}:   

\begin{eqnarray}
S_{F}^{opt}\left(\epsilon + \omega_{q}, \epsilon \right) = S(\epsilon + \omega_{q}) \frac{ \hbar\Omega }{\Delta }  \frac{  \sin(\pi  \omega_{q})  }{\pi \omega_{q}  }\, e^{ -i\pi \omega_{q} }  
\nonumber \\
\times \left\{ \delta(\omega_{q})   -    \frac{ \frac{ \exp \left( i 2\pi \omega_{q}\frac{ t_{-} }{\tau }   \right)  }{\left( 1 - \omega_{q} \right) } +  \frac{  \exp \left( i 2\pi \omega_{q}\frac{ t_{+} }{\tau }   \right)  }{\left( 1 + \omega_{q} \right) } }{ \rho^{*}(\epsilon + \omega_{q}) \rho(\epsilon) }  \right\}\,, 
\label{45} 
\end{eqnarray}
\begin{eqnarray}
S_{F}^{opt\,*} \left( \epsilon + \bar\omega + \omega_{q}  ,\epsilon + \bar\omega + \omega_{m}  \right) =  S^{*}(\epsilon + \bar\omega + \omega_{q} )
\nonumber \\
\nonumber \\
\times \frac{ \hbar\Omega }{\Delta }  \frac{  \sin(\pi  [ \omega_{m} - \omega_{q}])  }{\pi [ \omega_{m} - \omega_{q}]  }\, e^{ -i\pi [ \omega_{m} - \omega_{q}] }  \Bigg\{ \delta( \omega_{m} - \omega_{q})  
\nonumber \\
\left.
 -    \frac{ \frac{ \exp \left( i 2\pi [ \omega_{m} - \omega_{q}]\frac{ t_{-} }{\tau }   \right)  }{\left( 1 +  [\omega_{m} - \omega_{q}] \right) } +  \frac{  \exp \left( i 2\pi [ \omega_{m} - \omega_{q}]\frac{ t_{+} }{\tau }   \right)  }{\left( 1 - [ \omega_{m} - \omega_{q}] \right) } }{ \rho(\epsilon + \bar\omega + \omega_{q} ) \rho^{*}(\epsilon + \bar\omega + \omega_{m} ) }  \right\}\,,
\label{46} 
\end{eqnarray}
\ \\
\noindent
where we use the short notation $\rho(\epsilon)$ introduced in Eq.~(\ref{19}) and rewrite the stationary scattering amplitude, Eq.~(\ref{16-02}), as follows:

\begin{eqnarray}
S(\epsilon) = e^{ i(2\pi \epsilon + \theta_{r}) } \frac{\rho^{*}(\epsilon) }{\rho(\epsilon) } \,.
\label{strho}
\end{eqnarray}
\ \\
\noindent 
Here we took into account that under optimal conditions the phase is, $\phi^{opt}(\epsilon) = \pi + 2\pi \epsilon$, see Eq.~(\ref{14}).  

Using Eqs.~(\ref{45}) and (\ref{46}) we calculate  

\begin{eqnarray}
\Pi_{m}(\epsilon,\bar\omega) = e^{-i \pi \omega_{m}} \int\limits_{-\infty}^\infty  d\omega_{q} \,S(\epsilon + \omega_{q})\, S^{*}(\epsilon + \bar\omega + \omega_{q} )  
\nonumber \\
\times
\frac{  \sin(\pi  \omega_{q})  }{\pi \omega_{q}  }\,
 \frac{  \sin(\pi  [ \omega_{m} - \omega_{q}])  }{\pi [ \omega_{m} - \omega_{q}]  } \Bigg\{ \delta(\omega_{q}) 
\nonumber \\
\times  
 \left.
  -    \frac{ \frac{ \exp \left( i 2\pi \omega_{q}\frac{ t_{-} }{\tau }   \right)  }{\left( 1 - \omega_{q} \right) } +  \frac{  \exp \left( i 2\pi \omega_{q}\frac{ t_{+} }{\tau }   \right)  }{\left( 1 + \omega_{q} \right) } }{ \rho^{*}(\epsilon + \omega_{q}) \rho(\epsilon) }  \right\} \Bigg\{ \delta( \omega_{m} - \omega_{q}) 
\nonumber \\
\times  
 \left.
 -    \frac{ \frac{ \exp \left( i 2\pi [ \omega_{m} - \omega_{q}]\frac{ t_{-} }{\tau }   \right)  }{\left( 1 +  [\omega_{m} - \omega_{q}] \right) } +  \frac{  \exp \left( i 2\pi [ \omega_{m} - \omega_{q}]\frac{ t_{+} }{\tau }   \right)  }{\left( 1 - [ \omega_{m} - \omega_{q}] \right) } }{ \rho(\epsilon + \bar\omega + \omega_{q} ) \rho^{*}(\epsilon + \bar\omega + \omega_{m} ) }  \right\} . 
 \quad 
\label{47} 
\end{eqnarray}
\ \\
\noindent
After opening the brackets we use Eq.~(\ref{strho}) and get 

\begin{eqnarray}
\Pi_{m}(\epsilon,\bar\omega) = e^{-i \pi \omega_{m}} \,S(\epsilon)\, S^{*}(\epsilon + \bar\omega)   
\nonumber \\
\times 
\left\{ \delta\left( \omega_{m} \right) + \frac{ \sum_{j=1}^{3} {\cal A}_{j}  }{\rho(\epsilon + \bar\omega ) \rho^{*}(\epsilon + \bar\omega + \omega_{m} )  } \right\} ,
\label{48} 
\end{eqnarray}
\begin{eqnarray}
{\cal A}_{1} = -\, \frac{  \sin(\pi  \omega_{m})  }{\pi \omega_{m}  }\, \left\{ \frac{ e^{ i 2\pi \omega_{m}  \frac{ t_{-} }{\tau }  }  }{\left( 1 +  \omega_{m}  \right) } +  \frac{ e^{ i 2\pi \omega_{m} \frac{ t_{+} }{\tau }   }  }{\left( 1 -  \omega_{m}  \right) } \right\}\,,
\label{49}
\end{eqnarray}
\begin{eqnarray}
{\cal A}_{2} &=& -\, \frac{  \sin(\pi  \omega_{m})  }{\pi \omega_{m}  }\, \left\{ \frac{ e^{ i 2\pi \omega_{m}  \frac{ t_{-} }{\tau }   }  }{\left( 1 -  \omega_{m}  \right) } +  \frac{  e^{ i 2\pi \omega_{m} \frac{ t_{+} }{\tau }   }  }{\left( 1 +  \omega_{m}  \right) } \right\}
\nonumber \\
&&\times \frac{ \rho(\epsilon + \bar\omega + \omega_{m}) \rho^{*}(\epsilon + \bar\omega) }{\rho(\epsilon +  \omega_{m}) \rho^{*}(\epsilon )  } \,,
\label{50} 
\end{eqnarray}
\begin{eqnarray}
{\cal A}_{3} =  \frac{ \rho^{*}(\epsilon + \bar\omega) }{ \rho^{*}(\epsilon ) } \int\limits_{-\infty}^\infty  \frac{ d\omega_{q}  }{\rho(\epsilon + \omega_{q})\rho^{*}(\epsilon + \bar\omega + \omega_{q}) } \, 
\nonumber \\
\times  
\frac{  \sin(\pi  \omega_{q})  }{\pi \omega_{q}  }\,
 \frac{  \sin(\pi  [ \omega_{q} - \omega_{m}])  }{\pi [ \omega_{q} - \omega_{m}]  }
\nonumber \\
\times  
\left\{ \zeta_{q} +    {\cal B}_{1} e^{ i 2\pi \omega_{m}\frac{ t_{-} }{\tau }   }   +  {\cal B}_{2}  e^{ i 2\pi \omega_{m}\frac{ t_{+} }{\tau }   }   \right\} ,
\label{51} 
\end{eqnarray}
where
\begin{eqnarray}
{\cal B}_{1} = \frac{  1 }{\left( \omega_{q} - 1 \right) \left(  \omega_{q} - \omega_{m}  - 1 \right) } \,, 
\nonumber \\
{\cal B}_{2} = \frac{  1 }{\left( \omega_{q} + 1 \right)\left( \omega_{q} - \omega_{m} + 1 \right) } \,.
\label{52} 
\end{eqnarray}
\ \\
\noindent
The quantity  $\zeta_{q}$ contains a factor oscillating fast in $\omega_{q}$ and, therefore, under the condition set out in Eq.~(\ref{08}) $\zeta_{q}$ vanishes upon integration over $\omega_{q}$.

\subsection{Zero-frequency limit}
\label{zfl}

At $\omega = 0$ we calculate ${\cal A}_{3}$, Eq.~(\ref{51}), in the same way as we calculated Eq.~(\ref{24}): 
We use Eq.~(\ref{26}) with the following modifications: $\omega_{p} \to \omega_{q}$,  $\omega_{n} = 0$. 
In addition we change the sign of $\omega_{m}$. 
The result is, ${\cal A}_{3} = - ({\cal A}_{1} + {\cal A}_{2})$.   
Then we use Eq.~(\ref{48}) and find the noise power, Eq.~(\ref{39-1}), to be zero as it is expected.

\subsection{Finite-frequency noise}
\label{ffn1}

In the small transmission limit, $\bar T\ll 1$, we can calculate ${\cal A}_{3}$, Eq.~(\ref{51}), analytically. 
For this purpose we represent $1/\rho(\epsilon)$, Eq.~(\ref{19}), as the sum of the Breit-Wigner resonances. 
For optimal operating conditions, Eq.~(\ref{14}), it reads, 

\begin{eqnarray}
\frac{1}{\rho(x) } &=& \sum\limits_{a=-\infty}^{\infty} \frac{i \sqrt{g/\pi} }{ x + 0.5 - a + i g }\,,  
\label{53} 
\end{eqnarray}
\ \\
\noindent
where we introduce a short notion $g \equiv \bar\delta/\Delta = \bar T/(4\pi)$. 
Since the width $\bar\delta$ of peaks is small and the adjacent peaks (separated by the level spacing $\Delta$) do not overlap, we can keep only the pair products of closest peaks in the product of $\rho-$functions entering the integral in Eq.~(\ref{51}):

\begin{eqnarray}
\frac{1}{\rho(\epsilon + \omega_{q})\rho^{*}(\epsilon+\bar\omega + \omega_{q} )} \approx 
\nonumber \\
 \sum\limits_{a=-\infty}^{\infty}\!\! \frac{g/\pi }{ \left( \eta - a + i g \right)\left( \eta  + \bar\omega  - a - i g \right) } \,.
\label{54} 
\end{eqnarray}
\ \\
\noindent
where we use a short notation $\eta = \epsilon  + \omega_{q}+ 0.5$. 
Notice for frequencies close  to $\Delta$ we have to re-pair peaks.
This can be done with the help of the shift $\bar\omega \to \bar\omega - 1$. 
 
Furthermore we make a shift, $\omega_{q} \to \omega_{q} - \epsilon \pm 0.5 + a$ (the upper/lower sign is used in the term with ${\cal B}_{1}$/${\cal B}_{2}$), and get

\begin{eqnarray}
{\cal A}_{3} =  \frac{ \rho^{*}(\epsilon + \bar\omega) }{ \rho^{*}(\epsilon ) } \frac{g }{\pi } \int\limits_{-\infty}^\infty  \frac{ d\omega_{q} \Xi \left\{  e^{ i 2\pi \omega_{m}\frac{ t_{-} }{\tau }   }   +    e^{ i 2\pi \omega_{m}\frac{ t_{+} }{\tau }   }   \right\}   }{  \left( \omega_{q} + i g \right)\left(  \bar\omega + \omega_{q}  - i g \right) } \,, 
\nonumber \\
\Xi = \sum\limits_{a=-\infty}^{\infty}
\frac{  \cos(\pi  y_{q} )  \cos(\pi  [ y_{q}   - \omega_{m}])   }{\pi^{2} \left\{(a  + y_{q} )^{2} - \frac{1}{4} \right\}\left\{(a  + y_{q} - \omega_{m})^{2} - \frac{1}{4} \right\}  } 
 \,,
\nonumber \\
\label{55} 
\end{eqnarray}
\ \\
\noindent
where we introduce $y_{q} = \omega_{q} - \epsilon$.
We use Eq.~(\ref{30}) with $x = - \omega_{m}$ and find that $\Xi$ is independent of $\omega_{q}$,

\begin{eqnarray}
\Xi = \frac{2\sin\left( \pi \omega_{m}  \right) }{\pi \omega_{m}  \left( 1 - \omega_{m}^{2} \right) } \, .
\label{xi01}
\end{eqnarray}
\ \\
\noindent
The integration in Eq.~(\ref{55}) becomes trivial and we obtain

\begin{equation}
{\cal A}_{3} =  \frac{ \rho^{*}(\epsilon + \bar\omega) }{ \rho^{*}(\epsilon ) } \frac{2\sin\left( \pi \omega_{m} \right) }{\pi \omega_{m} \left( 1 - \omega_{m}^{2} \right) } \frac{  e^{ i 2\pi \omega_{m}\frac{ t_{-} }{\tau }   }   +    e^{ i 2\pi \omega_{m}\frac{ t_{+} }{\tau }   }   }{1 + i \varpi } \,, 
\label{56}
\end{equation}
\ \\
\noindent
where $\varpi = \bar\omega/(2g) \equiv \omega\tau_{D}$. 

Substituting all ${\cal A}_{j}$'s into Eq.~(\ref{48}) we find

\begin{eqnarray}
\Pi_{m}(\epsilon,\bar\omega) = e^{-i \pi \omega_{m}} \,S(\epsilon)\, S^{*}(\epsilon + \bar\omega)   
\nonumber \\
\times 
\left\{\! \delta\left( \omega_{m} \right) + \frac{\sin(\pi\omega_{m})  \left(   {\cal C}_{1} e^{ i 2\pi \omega_{m}\frac{ t_{-} }{\tau }   }   +  {\cal C}_{2}  e^{ i 2\pi \omega_{m}\frac{ t_{+} }{\tau }   }  \right)  }{\pi\omega_{m} \left( 1 -  \omega_{m}^{2}  \right) } \!   \right\} ,
\nonumber \\
\label{57} 
\end{eqnarray}
with
\begin{eqnarray}
{\cal C}_{1} = \frac{ 2 }{ 1+  i \varpi } \frac{\rho^{*}\left( \epsilon + \bar\omega  \right) }{\rho\left( \epsilon + \bar\omega  \right) \rho^{*}\left( \epsilon \right) \rho^{*}\left( \epsilon + \bar\omega + \omega_{m} \right) }
\nonumber \\
+  \frac{\omega_{m} -1 }{\rho\left( \epsilon + \bar\omega \right) \rho^{*}\left( \epsilon + \bar\omega + \omega_{m} \right) }
\nonumber \\
\nonumber \\
- \frac{\omega_{m} +1 }{\rho^{*}(\epsilon ) \rho(\epsilon +  \omega_{m})  }\frac{ \rho^{*}(\epsilon + \bar\omega)\rho(\epsilon + \bar\omega + \omega_{m})  }{ \rho\left( \epsilon + \bar\omega \right) \rho^{*}\left( \epsilon + \bar\omega + \omega_{m} \right) }
\,,
\label{58} 
\end{eqnarray}
\begin{eqnarray}
{\cal C}_{2} = \frac{ 2 }{ 1+  i \varpi } \frac{\rho^{*}\left( \epsilon + \bar\omega  \right) }{\rho\left( \epsilon + \bar\omega  \right)\rho^{*}\left( \epsilon \right) \rho^{*}\left( \epsilon + \bar\omega + \omega_{m} \right) }
\nonumber \\
-  \frac{\omega_{m} +1 }{\rho\left( \epsilon + \bar\omega \right) \rho^{*}\left( \epsilon + \bar\omega + \omega_{m} \right) }
\nonumber \\
\nonumber \\
+ \frac{\omega_{m} -1 }{\rho^{*}(\epsilon ) \rho(\epsilon +  \omega_{m})  }\frac{ \rho^{*}(\epsilon + \bar\omega)\rho(\epsilon + \bar\omega + \omega_{m})  }{ \rho\left( \epsilon + \bar\omega \right) \rho^{*}\left( \epsilon + \bar\omega + \omega_{m} \right) }
\,.
\label{59} 
\end{eqnarray}
\ \\
\noindent
As the next step we square $\Pi_{m}(\epsilon,\bar\omega)$, Eq.~(\ref{57}), and calculate

\begin{eqnarray}
\left| \Pi_{m}(\epsilon,\bar\omega)\right|^{2} &=& \delta^{2}\left( \omega_{m} \right)  +  2 \delta\left( \omega_{m} \right) {\rm Re} \left [ {\cal C}_{1} + {\cal C}_{2} \right]  
\nonumber \\
&&+ \frac{\sin^{2}(\pi\omega_{m}) \left( |{\cal C}_{1}|^{2} +  |{\cal C}_{2}|^{2} \right) }{\pi^{2}\omega_{m}^{2} \left( 1 - \omega_{m}^{2}  \right)^{2} } + \zeta_{m} 
\,.
\quad \quad
\label{60} 
\end{eqnarray}
\ \\
\noindent
Here $\zeta_{m}$ is oscillating fast in $\omega_{m}$ and thus it does not contribute to the noise power, Eq.~(\ref{39-1}). 

Next we calculate the necessary quantities,

\begin{eqnarray}
\left | {\cal C}_{1} \right |^{2} = \frac{(\omega_{m} + 1)^{2}  }{|\rho(\epsilon)|^{2} |\rho(\epsilon +  \omega_{m})|^{2} } 
\nonumber \\ 
+ \frac{(\omega_{m} -1)^{2} }{|\rho(\epsilon + \bar\omega)|^{2} |\rho(\epsilon + \bar\omega + \omega_{m})|^{2} }  
\nonumber \\
\nonumber \\ 
 + \frac{4 }{1 + \varpi^{2} } \frac{1 }{|\rho(\epsilon)|^{2} |\rho(\epsilon + \bar\omega +\omega_{m})|^{2} } 
\nonumber \\
\nonumber \\ 
 - 2 {\rm Re} \frac{\omega_{m}^{2} - 1 }{ \rho\left( \epsilon \right)  \rho^{*}\left( \epsilon  + \omega_{m} \right) \rho^{*}\left( \epsilon + \bar\omega \right) \rho\left( \epsilon + \bar\omega + \omega_{m} \right)   }
\nonumber \\ 
\nonumber \\
 - 2 {\rm Re} \frac{2(\omega_{m} + 1) }{(1 + i \varpi )|\rho(\epsilon)|^{2}\rho\left( \epsilon + \bar\omega + \omega_{m} \right) \rho^{*}\left( \epsilon +  \omega_{m} \right) }
\nonumber \\ 
\nonumber \\
 + 2 {\rm Re} \frac{2(\omega_{m} - 1) }{(1 + i\varpi ) |\rho(\epsilon + \bar\omega + \omega_{m})|^{2}\rho^{*}\left( \epsilon  \right) \rho\left( \epsilon +  \bar\omega \right) }
\,, 
\nonumber \\
\label{61} 
\end{eqnarray}
\begin{eqnarray}
\left | {\cal C}_{2} \right |^{2} = \frac{(\omega_{m} - 1)^{2}  }{|\rho(\epsilon)|^{2} |\rho(\epsilon +  \omega_{m})|^{2} } 
\nonumber \\ 
+ \frac{(\omega_{m} +1)^{2} }{|\rho(\epsilon + \bar\omega)|^{2} |\rho(\epsilon + \bar\omega + \omega_{m})|^{2} }  
\nonumber \\ 
\nonumber \\
 + \frac{4 }{1 + \varpi^{2} } \frac{1 }{|\rho(\epsilon)|^{2} |\rho(\epsilon + \bar\omega +\omega_{m})|^{2} } 
\nonumber \\ 
\nonumber \\
 - 2 {\rm Re} \frac{\omega_{m}^{2} - 1 }{ \rho\left( \epsilon \right) \rho^{*}\left( \epsilon  + \omega_{m} \right)\rho^{*}\left( \epsilon + \bar\omega \right) \rho\left( \epsilon + \bar\omega + \omega_{m} \right)   }
\nonumber \\ 
\nonumber \\
 + 2 {\rm Re} \frac{2(\omega_{m} - 1) }{(1 + i \varpi )|\rho(\epsilon)|^{2}\rho\left( \epsilon + \bar\omega + \omega_{m} \right) \rho^{*}\left( \epsilon +  \omega_{m} \right) }
\nonumber \\ 
\nonumber \\
 - 2 {\rm Re} \frac{2(\omega_{m} + 1) }{(1 + i \varpi ) |\rho(\epsilon + \bar\omega + \omega_{m})|^{2}\rho^{*}\left( \epsilon  \right) \rho\left( \epsilon +  \bar\omega \right) }
\,. 
\nonumber \\
\label{62} 
\end{eqnarray}
\ \\
\noindent
We sum up these equations

\begin{eqnarray}
\left | {\cal C}_{1} \right |^{2} + \left | {\cal C}_{2} \right |^{2} = \frac{2(\omega_{m}^{2} + 1)  }{|\rho(\epsilon)|^{2} |\rho(\epsilon +  \omega_{m})|^{2} } 
\nonumber \\
+ \frac{2(\omega_{m}^{2} + 1) }{|\rho(\epsilon + \bar\omega)|^{2} |\rho(\epsilon + \bar\omega + \omega_{m})|^{2} }  
\nonumber \\
\nonumber \\
 + \frac{8 }{1 + \varpi^{2} } \frac{1 }{|\rho(\epsilon)|^{2} |\rho(\epsilon + \bar\omega +\omega_{m})|^{2} } 
\nonumber \\
\nonumber \\
 - 4 {\rm Re} \frac{\omega_{m}^{2} - 1 }{\rho\left( \epsilon \right) \rho^{*}\left( \epsilon + \bar\omega \right) \rho^{*}\left( \epsilon  + \omega_{m} \right) \rho\left( \epsilon + \bar\omega + \omega_{m} \right)   }
\nonumber \\
\nonumber \\
 - 8 {\rm Re} \frac{(1 - i\omega\tau_{D}) }{(1 +\varpi^{2})|\rho(\epsilon)|^{2} \rho^{*}\left( \epsilon +  \omega_{m} \right) \rho\left( \epsilon + \bar\omega + \omega_{m} \right) }
\nonumber \\
\nonumber \\
 - 8 {\rm Re} \frac{(1 - i\omega\tau_{D}) }{(1 + \varpi^{2}) |\rho(\epsilon + \bar\omega + \omega_{m})|^{2}\rho^{*}\left( \epsilon  \right) \rho\left( \epsilon +  \bar\omega \right) }
\,. 
\nonumber \\
\label{63} 
\end{eqnarray}
\ \\
\noindent
Also we need the following

\begin{eqnarray}
\delta\left( \omega_{m} \right) {\rm Re} \left [ {\cal C}_{1} + {\cal C}_{2} \right] = {\rm Re} \left [  \frac{ 4 }{ 1+  i \varpi } \frac{1}{\rho\left( \epsilon + \bar\omega  \right) \rho^{*}\left( \epsilon \right) } 
\right.
\nonumber \\
\left.
- \frac{2 }{ |\rho(\epsilon + \bar\omega) |^{2} }
- \frac{2 }{ |\rho(\epsilon) |^{2} }
\right] \delta\left( \omega_{m} \right) \,.
\quad 
\label{64} 
\end{eqnarray}
\ \\
\noindent
We use Eqs.~(\ref{63}) and (\ref{64}) in Eq.~(\ref{60}) and represent the noise power, Eq.~(\ref{39-1}), as the sum of seven terms, ${\cal P}^{ex}= \sum_{j=1}^{7} {\cal P}_{j}(\omega)$. 
The first six terms originate from six terms in Eq.~(\ref{63}) and the seventh one results from Eq.~(\ref{64}). 
In the first three terms we integrate over both $\epsilon$ and $\omega_{m}$ using the series of the Breit-Wigner peaks, see, Eq.~(\ref{53}).
Ignoring small terms of order $g = \bar T /(4\pi) \ll 1$ we treat these peaks as delta-function peaks,  

\begin{eqnarray}
\frac{1}{|\rho(x) |^{2}} &\approx& \sum\limits_{a=-\infty}^{\infty} \delta\left( x + 0.5 - a \right)\,,  
\label{65} 
\end{eqnarray} 
\ \\
\noindent
and calculate,

\begin{eqnarray}
{\cal P}_{1}(\bar\omega) = \frac{{\rm e}^{2} \Omega }{2\pi} \sum\limits_{a=-\infty}^{\infty} 
\left\{   
2F ( {a - 0.5 , a-0.5 + \bar\omega } )  
\right. 
\nonumber \\
\left.
+  F ( {a - 0.5 , a+0.5 + \bar\omega } ) + F ( {a + 0.5 , a-0.5 + \bar\omega } ) 
\right\}
\,,
\nonumber \\
\label{66} 
\end{eqnarray}
\begin{eqnarray}
{\cal P}_{2}(\bar\omega) = \frac{{\rm e}^{2} \Omega }{2\pi} \sum\limits_{a=-\infty}^{\infty} 
\left\{   
2F ( a-0.5  ,a - 0.5 - \bar\omega  )  
\right. 
\nonumber \\
\left.
+  F ( a+0.5, a - 0.5 - \bar\omega   ) + F ( a - 0.5, a + 0.5 - \bar\omega   ) 
\right\}
\,,
\nonumber \\
\label{67} 
\end{eqnarray}
\begin{eqnarray}
{\cal P}_{3}(\bar\omega) = \frac{{\rm e}^{2} \Omega }{2\pi} 
\frac{8 }{1 + \varpi^{2} }
\sum\limits_{a=-\infty}^{\infty} 
\sum\limits_{b=-\infty}^{\infty} 
\nonumber \\
F ( a - 0.5 , a + b - 0.5  )  
\frac{\sin^{2}(\pi \bar\omega )  }{\pi^{2}[b  - \bar\omega]^{2} \left( 1 - [b  - \bar\omega]^{2}  \right)^{2} }\,.
\nonumber \\
\label{68} 
\end{eqnarray} 
\ \\
\noindent
Notice in Eqs.~(\ref{66}) and (\ref{67}) we keep only the terms corresponding to $\omega_{m} = 0\,, \pm 1$, since others vanish due to the factor $\sin^{2}(\pi \omega_{m})$, see Eq.~(\ref{60}).  
In the term with $\omega_{m} = - 1$ we made a shift $a \to a + 1$ in the infinite sum over $a$. 
In addition we used the following symmetries [see Eqs.~(\ref{002}) and (\ref{contfermi})]

\begin{eqnarray}
F(\epsilon_{1},\epsilon_{2}) = F(\epsilon_{2},\epsilon_{1})\,, \ \ F(\epsilon_{1},\epsilon_{2}) = F(-\epsilon_{1},-\epsilon_{2})\,.
\label{fsym}
\end{eqnarray}

To calculate the remaining contributions we use the expansions similar to  Eq.~(\ref{54}) and proceed as follows:

\begin{eqnarray}
\int  \frac{ d\epsilon \, \Psi(\epsilon) }{\rho^{*}\left( \epsilon \right) \rho\left( \epsilon + \bar\omega  \right)  } = \int  \frac{dx \, g/\pi}{(x - ig)( x + \bar\omega + ig) } 
\nonumber \\
\times  \sum\limits_{a = - \infty}^{\infty} \frac{ \Psi(a - 0.5) + \Psi(a - 0.5 - \omega) }{2 } 
\nonumber \\
=  \frac{1 }{2 }\sum\limits_{a = - \infty}^{\infty} \frac{ \Psi(a - 0.5) + \Psi(a - 0.5 - \omega)  }{1 - i \varpi }
\,,
\label{69} 
\end{eqnarray}
\ \\
\noindent
where $\Psi(\epsilon)$ is a function varying slow on the scale of $g\ll 1$. 
Notice the poles of $1/\rho^{*}\left( \epsilon \right)$ are at $\epsilon^{p} = a - 0.5 + ig$ while those of $1/\rho\left( \epsilon + \bar\omega  \right)$ are at $\epsilon^{p} = a - 0.5 - \bar\omega - ig$.
In the argument of $\Psi$ we ignore the term $x \sim g$ compared to others. 
This allowed us to integrate over $x$. 

By analogy we use, 

\begin{eqnarray}
\iint   \frac{ d\epsilon d\omega_{m} \Psi(\epsilon,\omega_{m}) }{
\rho\left( \epsilon \right) 
\rho^{*}\left( \epsilon + \bar\omega \right) 
\rho^{*}\left( \epsilon  + \omega_{m} \right) 
\rho\left( \epsilon + \bar\omega + \omega_{m} \right)  
} = 
\nonumber \\
\iint    
\frac{ dx g/\pi}{(x + ig)( x + \bar\omega - ig) } 
\frac{ dy g/\pi}{(y - ig)( y + \bar\omega + ig) } 
\nonumber \\
\sum\limits_{a = - \infty}^{\infty} 
\sum\limits_{b = - \infty}^{\infty} 
\frac{\Psi(a - 0.5, b - a) + \Psi(a - 0.5 - \bar\omega, b - a) }{2 } 
\nonumber \\
=  
\sum\limits_{a = - \infty}^{\infty} 
\sum\limits_{b = - \infty}^{\infty} 
\frac{ \Psi(a - 0.5, b - a) + \Psi(a - 0.5 - \bar\omega, b - a)  }{ 2\left( 1 + i \varpi  \right) \left( 1 - i \varpi  \right) }
\,.
\nonumber \\
\label{69x1} 
\end{eqnarray}
\ \\
\noindent
Notice, we calculate $\Psi(\epsilon,\omega_{m})$ at the poles of $1/[\rho\left( \epsilon \right) \rho^{*}\left( \epsilon  + \omega_{m} \right) ]$ lying at $\epsilon^{p} = a - 0.5 - ig$ and $\omega_{m}^{p} = b  - a + 2ig$ and at the poles of $1/[\rho^{*}\left( \epsilon + \bar\omega \right) \rho\left( \epsilon + \bar\omega + \omega_{m} \right) ]$ lying at $\epsilon^{p} = a - 0.5 - \bar\omega + ig$ and $\omega_{m}^{p} = b  - a - 2ig$.

Thus the remaining contributions are following,

\begin{equation}
{\cal P}_{4}(\bar\omega) = \frac{{\rm e}^{2} \Omega }{2\pi} 
\frac{4 }{1 + \varpi^{2} }
\sum\limits_{a=-\infty}^{\infty} 
F ( a - 0.5 , a  - 0.5 + \bar\omega  )  
\,,
\label{70} 
\end{equation} 
\begin{eqnarray}
{\cal P}_{5}(\bar\omega) = {\cal P}_{6}(\bar\omega) =  - \, \frac{{\rm e}^{2} \Omega }{2\pi} 
\frac{1 }{1 + \varpi^{2} }
\sum\limits_{a=-\infty}^{\infty} 
\nonumber \\
\times
\Bigg\{   
4F ( a - 0.5 , a  - 0.5 + \bar\omega  )  
\nonumber \\
+  
F ( a - 0.5 , a+0.5 + \bar\omega  ) 
+ 
F ( a + 0.5 , a - 0.5 + \bar\omega  ) 
\nonumber \\
+ 
\sum\limits_{b=-\infty}^{\infty} 
\frac{4\sin^{2}(\pi \bar\omega ) F ( a - 0.5 , a + b - 0.5  )    }{\pi^{2}[b  - \bar\omega]^{2} \left( 1 - [b  - \bar\omega]^{2}  \right)^{2} }
\Bigg\}
\,.
\nonumber \\
\label{71} 
\end{eqnarray} 
\ \\
\noindent
Note that the double sums in ${\cal P}_{5}$ and ${\cal P}_{6}$ cancel out  the contribution ${\cal P}_{3}$. 

And finally the contribution due to Eq.~(\ref{64}) reads,

\begin{eqnarray}
{\cal P}_{7}(\bar\omega) = \frac{{\rm e}^{2} \Omega }{2\pi} \sum\limits_{a=-\infty}^{\infty} 
\left\{   
\frac{8 F ( a - 0.5, a - 0.5 + \bar\omega  ) }{1 + \varpi^{2} }  
\right. 
\nonumber \\
- 4  F (a - 0.5, a - 0.5 - \bar\omega) 
- \! 4 F (a - 0.5, a - 0.5 + \bar\omega) \!\! \bigg\}
\,,
\nonumber \\
\label{73} 
\end{eqnarray}
\ \\
\noindent
Summing up all the contributions and taking into account that $\varpi = \omega\tau_{D}$ we arrive at the equation (\ref{74}).

\section{Distribution function for single-particle excitations}
\label{dis-fun}

The distribution function for electrons, which are scattered off the dynamic single-channel sample, is calculated as follows, \cite{Moskalets:2002hu}

\begin{equation}
f_{out}(E) =  \sum_{q=-\infty}^{\infty} \left|S_{F}\left(E, E_{q}\right) \right|^{2} f\left(E_{q}\right)\,,
\label{d03}
\end{equation}
\ \\
\noindent
where $f(E)$ is the Fermi distribution function for electrons incoming from the reservoir. 
For zero temperature the Fermi function is $f(E) = \theta(\mu - E)$ and Eq.~(\ref{d03}) becomes

\begin{eqnarray}
f_{out}(E) =  \sum_{q=-\infty}^{\left [ (\mu - E)/(\hbar\Omega) \right]} \left|S_{F}\left(E, E_{q}\right) \right|^{2} \,.
\label{d04}
\end{eqnarray}
\ \\
\noindent
Here $\left [ X \right]$ stands for the integer part of $X$.

The dynamic scatterer excites an electron system and makes it non-equilibrium even if before scattering electrons were in equilibrium.   
The basic process leading to this non-equilibration is an energy exchange between the dynamic scatterer and electrons. 
Generally electrons absorb energy from the scatterer. 
As a result some unoccupied in equilibrium states become occupied and vice versa. 
The non-equilibrium excitations with $E>\mu$ are referred to as quasi-electrons and the ones with $E < \mu$ are referred to as holes.  
The distribution function for quasi-electrons can be defined as 

\begin{eqnarray}
f_{e}(E) = f_{out}(E) - f(E) \,,
\label{dfe}
\end{eqnarray}
and for holes as 
\begin{eqnarray}
f_{h}(E) = f(E) - f_{out}(E) \,.
\label{dfh}
\end{eqnarray}
\ \\
\noindent
Note, in the quantized emission regime instead of quasi-particle excitations of the Fermi sea in the electron waveguide it is more natural to speak about the particles, electrons and holes, emitted by the cavity.

\subsection{Adiabatic regime}
\label{dis-fun-ad}

In the adiabatic regime we use for the Floquet scattering matrix elements the following, $S_{F}(E, E_{q}) \approx S_{-q}(E)$, Eq.~(\ref{adfs}),  where the Fourier coefficients $S_{q}$ for the frozen scattering amplitude are given in Eq.~(\ref{07_nsc_10}).   
Using these equations we calculate the distribution function  at zero temperature. 
For $E > \mu$ the equation (\ref{dfe}) gives: 

\begin{eqnarray}
f_{e}^{ad}(E_{n})  &=&  2 \Omega \Gamma_{\tau} e^{-2\Omega\Gamma_{\tau}(n+1) } \,,
\label{a01}  
\end{eqnarray} 
\ \\
\noindent
where $n\hbar\Omega < E_{n} - \mu < (n+1)\hbar\Omega$ with $n = 0\,, 1\,, \dots$. 
The superscript ``$ad$'' stands for the adiabatic regime. 
With this distribution function we can calculate the number of emitted electrons per unit length of the waveguide:

\begin{eqnarray}
\delta n_{e}^{ad} &=& \int\limits_{0}^{\infty} \frac{d E}{h v_{\mu}} f_{e}^{ad}(E)  =  \frac{1}{{\cal T} v_{\mu}} \sum\limits_{n=0}^{\infty} f_{e}^{ad}(E_{n}) \nonumber \\
&\approx& \frac{1}{{\cal T} v_{\mu}} \, e^{-2\Omega\Gamma} \approx \frac{1}{{\cal T} v_{\mu}} \,. 
\label{a02}  
\end{eqnarray} 
\ \\
\noindent
So we have one electron emitted during each period ${\cal T}$, as expected.
Notice in above equation $v_{F}$ is the velocity of an electron with Fermi energy and $1/(h v_{\mu})$ is the density of states in the electron  waveguide at the Fermi energy. 

By analogy for $m\hbar\Omega < \mu - E_{m} < (m+1)\hbar\Omega$, with $m = 0\,, 1\,, \dots$, we calculate the hole distribution function, Eq.~(\ref{dfh}), in the adiabatic regime:

\begin{eqnarray}
f_{h}^{ad}(E_{m})  &=&  2 \Omega \Gamma_{\tau} e^{-2\Omega\Gamma_{\tau}(m+1) } \,.
\label{a02-1}  
\end{eqnarray} 

\subsection{Non-adiabatic regime}
\label{nar}

With notation introduced in Sec.~\ref{cfr} the distribution function $f_{out}$, Eq.~(\ref{d03}), reads  

\begin{equation}
f_{out}(\epsilon) = \frac{\Delta}{\hbar\Omega} \int\limits_{- \infty}^{\infty} d \omega_{q}\left| S_{F}(\epsilon, \epsilon + \omega_{q}) \right|^{2} f\left(\epsilon + \omega_{q} \right)\,.
\label{03-2}
\end{equation}
\ \\
\noindent
where we use $\epsilon = (E - \mu)/\Delta$ and $\omega_{q} = q\hbar\Omega/\Delta$. 

At zero temperature we have for quasi-electron excitations,  

\begin{equation}
f_{e}(\epsilon) = \frac{\Delta}{\hbar\Omega} \int\limits_{-\infty}^{-\epsilon} d \omega_{q} \left| S_{F}(\epsilon, \epsilon + \omega_{q}) \right|^{2} \,, \quad \epsilon > 0 \,,
\label{03-3}
\end{equation}
\ \\
\noindent
and for holes,  

\begin{equation}
f_{h}(\epsilon) = \frac{\Delta}{\hbar\Omega} \int\limits_{|\epsilon|}^{\infty} d \omega_{q} \left| S_{F}(\epsilon, \epsilon + \omega_{q}) \right|^{2} \,, \quad \epsilon < 0  \,,
\label{03-4}
\end{equation}
where we used  

\begin{eqnarray}
\int\limits_{-\infty}^{\infty} d\omega_{q} \left |  S_{F}(\epsilon, \epsilon + \omega_{q}) \right |^{2} =  \frac{\hbar\Omega }{\Delta } \,.
\label{uni-2-0} 
\end{eqnarray}
\ \\
\noindent
Above identity follows from the unitarity condition, Eq.~(\ref{22-1}), with $\omega_{n} = \omega_{m} = 0$ and $\omega_{p}$ replaced by $\omega_{q}$. 
We also used there 

\begin{eqnarray}
\delta(0) = \frac{\Delta }{ \hbar\Omega } \,,
\label{delta0}
\end{eqnarray}
\ \\
\noindent
which can be proven by re-introducing the sum over $q$ instead of the integral over $d\omega_{q}$.

\subsubsection{Optimal operating conditions}
\label{optimal1}

For the cavity driven by the periodic step potential and working under  optimal operating conditions,  the Floquet scattering amplitude is given in Eq.~(\ref{20}). 
Its square, we need to calculate Eq.~(\ref{03-2}), is the following, 

\begin{eqnarray}
\left |  S_{F}^{opt}\left(\epsilon, \epsilon + \omega_{q} \right) \right |^{2} =   \left(  \frac{ \hbar\Omega }{\Delta }  \right)^{2}  \frac{  \sin^{2}(\pi  \omega_{q})  }{\pi^{2} \omega_{q}^{2}  }  \nonumber \\
\times \left| \delta(\omega_{q})   +   \frac{ \frac{ \exp \left( i 2\pi \omega_{m}\frac{ t_{-} }{\tau }   \right)  }{\left( \omega_{q}  -1 \right) } -  \frac{  \exp \left( i 2\pi \omega_{q}\frac{ t_{+} }{\tau }   \right)  }{\left( \omega_{q}  +1 \right) } }{\rho^{*}(\epsilon) \rho(\epsilon - \omega_{q}) }  \right|^{2} \,. 
\label{1-sm-02} 
\end{eqnarray}
\ \\
\noindent
the superscript ``$opt$'' stands for optimal operating conditions. 

For our present purposes we can simplify above equation: Since $\omega_{q} = 0$ is out of the integration interval in both Eq.~(\ref{03-3}) and  Eq.~(\ref{03-4}), we can safely relax terms containing the factor $\delta(\omega_{q})$ and use,

\begin{eqnarray}
\left |  S_{F}^{opt}\left(\epsilon, \epsilon + \omega_{q} \right) \right |^{2}  \sim    \left(  \frac{ \hbar\Omega }{\Delta }  \right)^{2}  \frac{  \sin^{2}(\pi  \omega_{q})  }{\pi^{2} \omega_{q}^{2}  }  
\nonumber \\
\times \frac{1}{ |\rho(\epsilon)|^{2}  } \frac{1 }{|\rho(\epsilon- \omega_{q})|^{2}  }  \left\{   \frac{ 1  }{\left( \omega_{q}  -1 \right)^{2} } +  \frac{1 }{\left( \omega_{q}  +1 \right)^{2} } + \xi_{q} \right\} \,. 
\nonumber \\
\label{1-sm-01} 
\end{eqnarray}
\ \\
\noindent
Here $\xi_{q}$ contains terms that oscillate fast in $\omega_{q}$ and which we neglect upon integration over $\omega_{q}$. 

First, we use above equation and calculate the distribution function for electrons, Eq.~(\ref{03-3}). 
The integral over $\omega_{q}$ to the leading order in the small parameter $g = \bar\delta/\Delta \ll 1$ can be evaluated using Eq.~(\ref{65}) with $x = \epsilon - \omega_{q}$. 
The result is

\begin{eqnarray}
 f_{e}^{opt}(\epsilon) = \frac{\hbar\Omega}{\Delta} \frac{\cos^{2}(\pi \epsilon)}{\pi^{2} |\rho(\epsilon)|^{2}  } \, \zeta \,, \quad \epsilon > 0 \,,  \nonumber \\
\label{dfe01} 
\end{eqnarray}
where 
\begin{eqnarray}
 \zeta =  \sum\limits_{a = -\infty}^{0}  \left\{   \frac{ 1  }{\left( \left[ \epsilon - a \right]^{2} - \frac{1}{4} \right)^{2} } +   \frac{ 1  }{\left( \left[ \epsilon + 1 - a \right]^{2} - \frac{1}{4} \right)^{2} }  \right\} \nonumber \\
 \nonumber \\
= \frac{ 1  }{\left(  \epsilon^{2} - \frac{1}{4} \right)^{2} } + \sum\limits_{a = 1}^{\infty}     \frac{ 2  }{\left( \left[ \epsilon + a \right]^{2} - \frac{1}{4} \right)^{2} }  \,. 
\nonumber 
\end{eqnarray}
\ \\
\noindent
The main contribution to the factor $\zeta$ comes from $\epsilon \sim 0.5$ and $a = 0$. 
Thus we obtain 

\begin{eqnarray}
f_{e}^{opt}(\epsilon) &\approx& \hbar\Omega  \frac{\cos^{2}(\pi \epsilon) \nu^{opt}(\epsilon)}{\pi^{2}  \left(  \epsilon^{2} - \frac{1}{4} \right)^{2} }\,, \quad \epsilon > 0  \,. 
\label{dfe02} 
\end{eqnarray}
\ \\
\noindent
Here we introduce the frozen density of states (DOS) of the cavity working under optimal conditions, $\nu^{opt}(\epsilon) = 1/(|\rho(\epsilon)|^{2} \Delta)$, see Eqs.~(\ref{dos}), (\ref{06}) and (\ref{19}). 
Note, if the cavity is driven by the pulsed potential and the working conditions are optimal, the frozen DOS, $\nu^{opt}(\epsilon)$, is time-independent. 
Using Eq.~(\ref{53}) we find   

\begin{eqnarray}
\nu^{opt}(\epsilon) &=& \sum\limits_{a=-\infty}^{\infty} \frac{\bar\delta/\pi }{\Delta^{2} \left( \epsilon + 0.5 - a  \right)^{2} +  \bar\delta^{2} }\,,  
\label{dosBW}
\end{eqnarray}
\ \\
\noindent
where we re-introduce the level width $\bar\delta = g \Delta$.
The DOS peaks at $\epsilon_{a} = a - 0.5$. 
However, at this values the factor $\cos(\pi \epsilon)$ in Eq.~(\ref{dfe02}) vanishes. 
Therefore, at positive $\epsilon$ the leading contribution arises form the peak around $\epsilon_{1} = 0.5$ where the zero in the denominator compensate the zero in the numerator. 
Thus we arrive at the following equation for the distribution function of non-adiabatically emitted electrons ($\epsilon > 0$),

\begin{eqnarray}
f_{e}^{opt}(\epsilon) \approx \hbar\Omega  \frac{\cos^{2}(\pi \epsilon)  }{\pi^{2}  \left(  \epsilon^{2} - \frac{1}{4} \right)^{2} }\frac{\bar\delta/\pi }{\Delta^{2} \left( \epsilon - 0.5  \right)^{2} +  \bar\delta^{2} }  \,. 
\label{dfe02-1}
\end{eqnarray}
\ \\
\noindent
For practical purposes this equation can be simplified even further,  

\begin{eqnarray}
f_{e}^{opt}(\epsilon) \approx \hbar\Omega  \frac{\bar\delta/\pi }{\Delta^{2} \left( \epsilon - 0.5  \right)^{2} +  \bar\delta^{2} }\,, \quad 0 < \epsilon < 1 \,, 
\label{dfe02-2}
\end{eqnarray}
\ \\
\noindent
where we restrict $\epsilon = (E - \mu)/\Delta$ from above  (by one level spacing) to keep the same mean energy and its variance calculated with the help of either Eq.~(\ref{dfe02-1}) or Eq.~(\ref{dfe02-2}).  

Using Eq.~(\ref{1-sm-01}) in Eq.~(\ref{03-4}), we calculate the distribution function for holes, $f_{h}^{opt}(\epsilon) = f_{e}^{opt}(-\epsilon)$.

\end{document}